\documentclass[a4paper,11pt]{article}
\pdfoutput=1
\usepackage{jcappub}
\usepackage[T1]{fontenc}
\usepackage{subfigure}
\usepackage{hhline}
\usepackage{xcolor}
\usepackage{aas_macros}

\newcommand{\be}{\begin{equation}}
\newcommand{\ee}{\end{equation}}
\newcommand{\bea}{\begin{eqnarray}}
\newcommand{\eea}{\end{eqnarray}}
\newcommand{\beq}{\begin{eqnarray}}
\newcommand{\eeq}{\end{eqnarray}}

\title{\boldmath Blue Loops, Cepheids, and Forays into Axions}

\author[a]{Kaleb Anderson,}
\author[b]{Thomas C. Gehrman,}
\author[c,d]{Pearl Sandick,}
\author[b,e]{~Kuver Sinha,}
\author[b]{Edward Walsh,}
\author[b]{Tao Xu}
\affiliation[a]{Department of Physics, Brown University, Providence, RI 02912, USA}
\affiliation[b]{Homer L. Dodge Department of Physics and Astronomy, University of Oklahoma, Norman, OK 73019, USA}
\affiliation[c]{Department of Physics and Astronomy, University of Utah, Salt Lake City, UT 84112, USA}
\affiliation[d]{Laboratoire Univers et Particules de Montpellier, CNRS \& Universit\'{e} de Montpellier, France}
\affiliation[e]{Department of Physics and Astronomy, Rice University, Houston, TX 77005, USA}

\emailAdd{kaleb\_anderson@brown.edu}
\emailAdd{thomas.gehrman@ou.edu}
\emailAdd{pearl.sandick@utah.edu}
\emailAdd{kuver.sinha@ou.edu}
\emailAdd{edward.a.walsh-1@ou.edu}
\emailAdd{tao.xu@ou.edu}

\abstract{The blue loop stage of intermediate mass stars has been called a ``magnifying glass'', where even seemingly small effects in prior stages of evolution, as well as assumptions about stellar composition, rotation, and convection, produce discernible changes. As such, blue loops, and especially the existence and properties of Cepheids, can serve as a laboratory where feebly connected Beyond Standard Model particles such as axions  can be gainfully studied. We undertake a careful study of the effects of these putative particles on the blue loop, paying close attention to the evolution of the core potential and the hydrogen profile. Our simulations, performed with \texttt{MESA}, place bounds on the axion-photon coupling using the galactic Cepheid S Mus, with dynamically-determined mass of $6 M_\odot$, as a benchmark. The effects of varying convective overshoot on the core potential and hydrogen profile, and the ensuing changes in the axion constraints, are carefully studied.  Along the way, we explore the ``mirror principle'' induced by the hydrogen burning shell and contrast our results with those existing in the literature. Less conservative (but more stringent) bounds on the axion-photon coupling are given for a $9 M_\odot$ model, which is the heaviest that can be simulated if overshoot is incorporated, and tentative projections are given for a $12 M_\odot$ model, which is approximately the heaviest tail of the  mass distribution of galactic Cepheids determined by pulsation models using Gaia DR2. Our main message is that the reliable simulation and observation (ideally, through dynamical mass determination) of massive Cepheids constitutes an important frontier in axion searches, challenges in modeling uncertainties in the microphysics  of the blue loop stage notwithstanding.
}

\begin{document}
\maketitle
\flushbottom

\section{Introduction}
\label{sec:intro}

The idea of probing Beyond Standard Model (BSM) physics with stellar objects has a 
 venerable history \cite{Raffelt:1996wa, Raffelt:1990yz}. The last ten years, in particular, have witnessed an explosion of activity in this area, as the community has increasingly focused on sub-GeV dark sectors \cite{Adams:2022pbo, Essig:2013lka}. Searches for such light BSM species have been undertaken in a host of systems: neutron stars, supernovae, white dwarfs, mergers, black holes, and various other astrophysical environments \cite{Baryakhtar:2022hbu}. The BSM species of choice has often been the axion, or more generally, axion-like particles; the current work, too, focuses on axions.\footnote{We will be interested in axion-like particles in this work, but will call them ``axions'' throughout.}

One way to organize such searches is to think of them as relating to the properties of individual stellar objects versus relating to the characteristics or behavior of entire stellar populations. Much of the recent work has been done in the context of  constraints on BSM species from \textit{individual} stellar objects, using the following cooling argument:  the object cannot emit BSM particles and therefore cool anomalously fast (typically compared to cooling induced by neutrino emission).  Constraints based on   \textit{population} studies, on the other hand, are argued as follows:  the emission of the BSM species and the attendant cooling alters the evolutionary trajectory of stellar objects, which is reflected at the population level. Such arguments can be potent if the BSM physics alters or eliminates a particularly long or significant portion of the evolution of a stellar population.

Stars like our Sun, as well as intermediate mass or even heavier stars, whose evolutionary histories rest on well-established theoretical models, immediately become a fertile ground for such enquiries. Various parts of the Hertzsprung-Russell (HR) diagram, which displays the evolutionary track of stars of various masses and provides a template on which stellar populations can be placed, become sites of investigation. A key target for BSM physics is evolutionary stages which feature a helium burning core, since  BSM particles are efficiently produced from such cores and the resulting loss of energy reduces the duration of the corresponding evolutionary phase. This reduction then gets reflected at the population level. Classic bounds on axions using this idea have come from the study of horizontal branch stars: the reduction in the duration of this phase lowers the ratio $R$ between the number of horizontal branch stars and red giant branch (RGB) stars in globular clusters \cite{Ayala:2014pea}. Recently, this has been extended to population ratios between horizontal branch stars and stars on the asymptotic giant branch (AGB) \cite{Dolan:2022kul}. 

A second target is constituted by the blue loop phase of intermediate mass stars. The blue loop stage on the HR diagram is the leftward movement of the evolutionary track, towards the hotter end of the HR diagram, followed by a subsequent rightward movement towards the AGB stage (cooler end). It is very sensitive to the physical properties of the star and to the various initial conditions inherited from previous phases, making it a long-standing laboratory to explore such effects (the classic text by \cite{Kippenhahn:2012qhp} describes it as ``a sort of magnifying glass, revealing relentlessly the faults of calculations of earlier phases'').  Energy loss due to the emission of BSM particles can inhibit the development of the blue loop, leading to constraints on such species. The blue loop stage acquires significance due to the existence of variable stars which fall within the instability strip. Cepheids, in particular, provide important potential sites where BSM physics may be explored, through their mass-luminosity (ML) and period-luminosity (PL) relations.  

The purpose of this paper is to perform a careful study of the effect of axion emission on the blue loop stage of intermediate mass stars, using a critical physical quantity as a diagnostic measure: the core potential \cite{Lauterborn:1971,Robertson:1972,Alongi:1991}. The core potential is defined in the following way
\be \label{corepot}
\phi_{c} \, = \, h \frac{M_c}{R_c} \,\,.
\ee
Here, $M_c$ and $R_c$ are the core mass and core radius, respectively, in solar units. The quantity $h$ is a function of the change in hydrogen abundance $\Delta X$ and the corresponding change in the mass profile $\Delta m$ when going from the core to the envelope across the hydrogen-burning shell: $h = h(\Delta X, \Delta m)$. The criterion for a blue loop, first introduced in \cite{Lauterborn:1971}, is as follows
\be \label{corepotcrit}
\phi_{c} \, < \, \phi_{\rm crit} \,\,\, \Rightarrow \,\,\, {\rm  \,\,blue\,\,loop,}
\ee
where $\phi_{\rm crit}$ is determined from simulations \cite{Kippenhahn:2012qhp}. It should be noted that the physical mechanisms behind the occurrence of a blue loop remain very much a topic of study in the astrophysics community. Nevertheless, through its dependence on the core radius and the width of the hydrogen shell, the core potential, via the criterion in Eq.~\ref{corepotcrit}, provides a relatively clean metric with which to diagnose emission due to BSM particles for benchmark values of rotation, composition, nuclear reaction rates, and convective mixing. It should be clarified right at the outset that the core potential, and in particular  its critical value, are only determined for a given model once the aforementioned physical properties are set to their benchmark values. The core potential, while providing a clean diagnostic for the onset of a loop, does not have predictive power for when or if a loop will develop.

There are two important aspects to criterion~\ref{corepotcrit} that can be highlighted. Firstly, it is the more massive stars that obey Eq.~\ref{corepotcrit} more robustly against changes in composition, convection, and rotation, and therefore it is these stars that have more robust blue loops. Secondly, we find that these same massive stars are much more likely to violate Eq.~\ref{corepotcrit} and lose their blue loops once BSM couplings are switched on.  The more massive stars with their hotter cores emit BSM species more efficiently (since such emission typically goes as the core temperature to some large power) and this emission inhibits the growth of the core, increases $\phi_c$, and ensures that Eq.~\ref{corepotcrit} is violated even for comparatively small BSM couplings.

We are able to perform controlled simulations of the core potential of up to $9 M_\odot$ models using the package \texttt{Modules for Experiments in Stellar Astrophysics (MESA)}~\cite{Paxton2010,Paxton2013,Paxton2015,Paxton2018,Paxton2019, Jermyn2023}.\footnote{More massive Cepheids can be simulated under the assumption of zero overshoot, as we will clarify later in the paper.} The axion coupling that just satisfies Eq.~\ref{corepotcrit} is obtained as a function of the Cepheid mass for simulated models up to $9 M_\odot$, and a fit provides projections for higher masses. This fit is provided in Fig.~\ref{fig:varyingmassconstraints} and can be used to obtain tentative constraints on the role of axions in curtailing blue loops in more massive stars, although such conclusions are not backed up by simulations and would be subject to greater uncertainties from convection models.

The main focus of our analysis are classical Cepheids in the Milky Way, with dynamically-determined masses.  The determination of masses by dynamical orbital methods that are unaffected by axion physics fixes the trajectory of these candidates on the HR diagram, for benchmark values of stellar input parameters. The presence of axions, however, does affect the evolution of the star as it approaches the blue loop stage, potentially halting the increase in temperature at $\phi_{c} \, \sim \, \phi_{\rm crit}$, prior to the star entering the instability strip.  The existence of the Cepheid variable can then be used to place a constraint on the axion coupling, which cannot be so large that the condition in Eq.~\ref{corepotcrit} is never satisfied. The most massive galactic Cepheid with a dynamically-determined mass, S Mus (with mass $6 M_\odot$) \cite{RemageEvans:2006pp} furnishes the most conservative constraint and is our benchmark result (the resulting constraint on the axion-photon coupling is an order of magnitude weaker than current limits from~\cite{Dolan:2022kul}). The potential future  observation and mass determination of a galactic Cepheid of  mass $9 M_\odot$ is at the limit of models that can currently be simulated with \texttt{MESA}, and would therefore furnish the strongest projected  constraint on axions if such an observation with dynamically determined mass were to be made.

Stronger constraints on axions from heavier Cepheids may very well be possible in the future. Masses of individual Cepheids determined by pulsation modeling using candidates in the Gaia DR2 galactic Cepheid database indicate a mass peak of around $\sim 5 M_\odot$, with data points as large as $\sim 12 M_\odot$ \cite{2020ApJ...898L...7M, 2019A&A...623A.117K}. Well-established ultra long-period Cepheids, such as OGLE-GD-CEP-1884, a galactic Cepheid with mass $7.47 M_\odot$  \cite{2024ApJ...965L..17S} determined by pulsation models, could also serve as benchmarks. A larger sample of Cepheid observations will also become available with the Gaia DR3 catalog \cite{2023A&A...674A..17R}. Stellar modeling indicates the possible existence of Cepheids of up to $20 M_\odot$, and, presumably, such a heavy Cepheid would yield the strongest constraint on the axion coupling. There are two primary reasons we do not venture to put axion limits from heavier Cepheids. Firstly, we are currently unable to simulate blue loops in models with masses larger than $9 M_\odot$, even in the absence of axions, for non-zero values of the overshoot (parametrized by $\alpha_{\rm ov}$ for alterations of the stellar interior structure by mixing chemical components beyond the convective boundary). Secondly, axions are expected to affect the pulsation models themselves. Nevertheless, we provide projections for a $12 M_\odot$ model, which, as expected, are very strong; we make no quantitative claims as to the robustness of those constraints. The variation of axion constraints as a function of stellar mass is discussed in Section~\ref{blueloopmass}, and the challenges of obtaining axion constraints for massive Cepheids is discussed in Section~\ref{sec: axionconstrmassive}, which also contains our main results. \textit{Our main message is that the detection and simulation of massive Cepheids is an important frontier in axion searches.} 

One of the biggest challenges confronting axion searches with stellar physics are the benchmark assumptions about the physical properties and composition of the star, and their possible impact in limiting sensitivity to BSM physics. For the case of Cepheids in the blue loop stage within the instability strip, these questions become especially important. There is a vast literature on the effect of nuclear reaction rates, rotation, metallicity, and convection  on the morphology of blue loops, and the addition of BSM physics would appear to add yet another uncertain aspect to a thicket of models. Among these astrophysical factors, we focus on the uncertainties caused by convection models, which are generally expected to be the most dominant, while keeping the metallicity at typical galactic benchmarks and excluding rotation altogether in this study. 
We undertake a careful study of the effect of overshoot on the core potential and probe how constraints on axions change with changing overshoot, as Eq.~\ref{corepotcrit} is violated. Limits on the axion coupling are displayed for various assumptions of the overshoot for the case of S Mus, and  a  $8 M_\odot$ model in Section~\ref{overshootuncert}.

In this context, we note the Cepheid mass discrepancy problem \cite{2023arXiv230712386G}:  while Cepheid masses determined from pulsation models  and binary orbital dynamics are largely in concordance with each other,  they are both  different from masses expected from stellar evolution. The mass discrepancy problem has been addressed by various methods, such as introducing overshoot, rotation, or pulsation-driven mass loss (we refer to \cite{2011A&A...529L...9N} for a review). Axions, we find, are unlikely to either alleviate or worsen this problem, since the effect of axions appears to be most relevant to the radius of the core and hence the core potential and the triggering of the blue loop. In other words, for a given stellar mass, axion emission does not move a blue loop vertically up or down, appreciably, in the HR diagram, but rather either allows or curtails the blue loop altogether.

Along the way, we explore various other aspects of the effect of BSM physics on the blue loop. Importantly, we note the physical origins of the blue loop come from the ``mirror principle'' \cite{Kippenhahn:2012qhp}: the fact that in the presence of a hydrogen burning shell, the two sides of the shell -- the envelope and the core -- develop oppositely, in the sense that when one contracts the other expands, and vice versa. During the blue loop, the core expands and the envelope contracts, drawing the star away from  the red giant phase with a large convective envelope towards becoming a compact blue giant. We explore the effect of BSM emission on the mirror principle in Fig.~\ref{fig:Radius}. In Fig.~\ref{fig:hydprofile} we explore the change in the hydrogen profile as the axion coupling is increased. The change in the core potential of a benchmark $6 M_\odot$ as the axion coupling is increased is depicted in Fig.~\ref{fig:corepot}; the change as the stellar mass is also increased is depicted in Fig.~\ref{fig:corepotmass}. 

Before proceeding, we pause to comment on previous work on BSM physics and blue loops. Our results are largely complementary to those obtained by \cite{2017A&A...605A.106C, Friedland:2012hj}\footnote{We note that \cite{Friedland:2012hj} had an error in their axion coupling computation, as already pointed out by \cite{2017A&A...605A.106C}.} -- in contrast to those studies, which focused on the relative reduction in the duration of the helium burning phase due to BSM emission, we focus on the core potential as a sharp diagnostic metric. The effects of overshoot were not considered by \cite{Friedland:2012hj}, who stressed the importance of incorporating the effect in future precision studies. Further points of distinction and contrast are detailed in the main text.

This paper is organized as follows. In Section~\ref{sec: ClassicalCepheid}, we provide an introduction to Cepheid variables and their corresponding simulations. Section~\ref{sec:axion} discusses the axion model and their production via the Primakoff effect in Cepheids. In Section~\ref{sec: blueloopaxion}, we explore the physics of the blue loop phase and the effects of axion energy loss. Constraints on the axion coupling and mass, based on the elimination of the blue loop, are presented in Section~\ref{sec: axionconstrmassive}. Finally, we conclude in Section~\ref{sec: conclusion}.

\section{Classical Cepheid Variables and Simulation Benchmarks}
\label{sec: ClassicalCepheid}

In this Section, we describe general properties of Cepheids and the various physical characteristics that affect their observational properties. 

Classical galactic Cepheids are young, intermediate mass, metal-rich, radially pulsating variable stars with typical ages ranging between $10-300$ Myr and masses between $4 - 12$ $M_{\odot}$. On the HR diagram, they occupy a narrow, nearly vertical band in the so-called instability strip. These luminous  giant variables  pulsate in the fundamental mode, the first overtone, second overtone harmonics, and multi-periodic modes.  

\begin{figure}[t]
  \centering
    \includegraphics[width=0.65\textwidth]{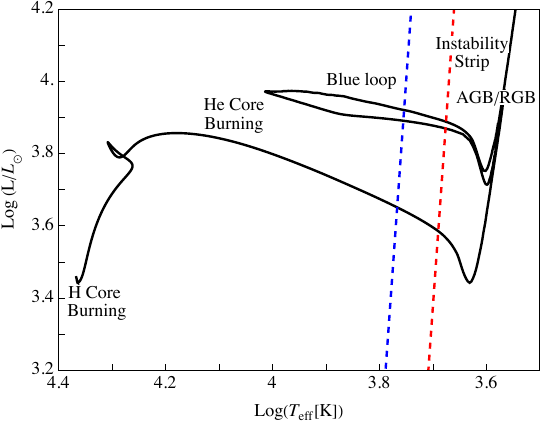}
  \caption{The evolutionary track of a non-rotating 8 $M_\odot$ \texttt{MESA} model. Included are important evolutionary stages of the  star and instability strips defined in \texttt{MESA}. The $L_\odot$ is the solar luminosity. The metallicity is set to $Z = 0.014$ and the overshoot to $\alpha_{\rm ov} = 0.0$. }
  \label{fig:evolution1}
\end{figure}

It is useful to recapitulate the main features of stellar evolution from the post-main sequence stage, leading up to the entry of Cepheid variables into the instability strip and their subsequent exit. Each model presented in our work starts from the zero age main sequence.  After the star exhausts the hydrogen in its core, it expands to become a red giant, initiating the RGB stage of its evolution. The expansion of the star, which has a temporarily inert helium core surrounded by a hydrogen-burning shell,  moves the star to the right (cooler temperature) in the HR diagram, constituting its first crossing through the instability strip. This crossing is rapid and typically not effective at constraining properties at the population level. The star crosses the red edge of the strip with its hydrogen shell still burning; thereafter, the helium core ignites, the star contracts and heats up, and proceeds to move to the left (hotter temperatures) on the HR diagram. The star crosses the instability strip for the second time (from the red edge), embarking on the ``blue loop'' phase. This helium burning phase lasts for a longer time than the first crossing (which is typically $10^3$ - $10^4$ years) and the star remains in the instability strip for a long time, eventually exhausting  the helium and moving right once again, to exit the strip. The star eventually climbs the AGB, with  the end product being either a white dwarf or a neutron star depending on the mass of the model. The main features of this evolutionary journey are shown for convenience in Fig.~\ref{fig:evolution1}.

The blue loop stage is the primary focus of our investigations. As mentioned in the Introduction, there is a vast literature on the physical attributes of the star and the various stellar processes that affect the progression of this stage. Nuclear reaction rates, rotation, metallicity, and convection have profound effects on the properties of blue loops, and hence the observational data pertaining to Cepheids.  
We use the 1-dimensional code \texttt{MESA} version \texttt{r23.05.1}.  We follow \cite{Grevesse1998} for composition and opacity settings in \texttt{MESA}. For wind mass loss, we use the ``Dutch'' wind scheme which follows \cite{Glebbeek2009,Vink2001,Nugis2000} with a scaling factor of 0.8. For mixing length theory, we follow \cite{Kuhfuss1986} with $\alpha_{\rm MLT}=1.6$ for our evolutionary models \cite{desomma2020,Bono2001}. We use the Ledoux criterion with semi-convection parameter $\alpha_{\rm semiconv}=0.1$ and the pp\_cno\_extras\_o18\_ne22.net for our nuclear reaction network \footnote{We note that according to \texttt{MESA}'s 2013 instrument paper, the semi-convection is treated as a time-dependent diffusive process following \cite{1983A&A...126..207L}.}. We terminate our models during a late stage of the AGB. Our models closely follow [32–35], with the metallicity set to
Z = 0.014 and a step overshoot over the hydrogen core is set to $\alpha_{ov}$ = 0.1 for the primary benchmarks. We set the rotation to zero in the current work, reserving the case of non-zero rotation for the future.

We make a few brief comments about the pipeline used in previous studies. The work of \cite{Friedland:2012hj} was the first (to our knowledge) to apply \texttt{MESA} to obtain constraints on axions from Cepheids. They used  \texttt{MESA} version 3794 to generate their models, assuming solar metallicity $Z = 0.02$ and zero overshoot $\alpha_{\rm ov} = 0.0$. Their study chose $\alpha_{\rm MLT} = 1.6$ and used  \cite{Grevesse1998} for the opacity tables and elemental mixture. On the other hand, \cite{2017A&A...605A.106C} used the \texttt{GENEVA} code, with $Z = 0.014$, $\alpha_{\rm ov} = 0.1$, and $\alpha_{\rm MLT} =1.6$. They used the elemental abundances of \cite{Asplund2005} and isotope mixture of \cite{Lodders2003}. Moreover, \cite{2017A&A...605A.106C} noted an issue with the calculation of  \cite{Friedland:2012hj}: the constant in front of Eq. (3) of \cite{Friedland:2012hj} is off by a factor of 10 ($c = 27.2$ ergs/g/s in \cite{Friedland:2012hj} instead of the correct value derived in Appendix A  of \cite{2017A&A...605A.106C}, which is $c = 283.16$ ergs/g/s).

Before proceeding, we make a few comments about the position and width of the instability strip, since it will play a vital role in what follows. The exact location of the instability strip depends on several factors, such as rotation and metallicity. For detailed modeling of these aspects, we refer to \cite{Anderson2014, Anderson_2016}.  In this paper, we define the instability strip according to Eq. (3) from \cite{Tammann_2003}. The width of the instability strip, given as $\Delta \log \, T_{\rm eff} = 0.08$ dex, is based on the distribution of Cepheid data from the Milky Way, Large Magellanic Cloud (LMC), and Small Magellanic Cloud (SMC) as described in \cite{Tammann_2003}.\footnote{We note that $\log$ in our work will have base 10, while the natural logarithm is denoted by $\ln$.} We express their Eq. (3) in terms of $\log L = a \log T_{\rm eff} + b$, where $a = - 20.83$ dex and $b = 81.2708$ dex. The red edge of the instability strip, as shown in our work, is determined by subtracting $\Delta \log \, T_{\rm eff} = 0.04$ dex from Eq. (3) of \cite{Tammann_2003}, while the blue edge is calculated by adding $\Delta \log \, T_{\rm eff} = 0.04$ dex to Eq. (3). It is important to note that axions may influence the location of the instability strip, potentially leading to observable consequences—an investigation we leave for future work \footnote{We note that there are three distinct temperatures that will be used in this work: $(i)$ $T_{\rm eff}$, the effective temperature of the surface of the star; $(ii)$ $T$, which denotes the temperature of the local cell under consideration during the relevant \texttt{MESA} simulation; and $(iii)$ $T_c$, which denotes the central temperature of the star.}.

\section{Axion Model and Primakoff Effect}
\label{sec:axion}

In this Section, we give details of our axion model, and how it is implemented in \texttt{MESA}. We use natural unit for the calculation of axion production in this section.

\subsection{Axion Production in Stars}
\label{sec:axionprimakoff}
We introduce the axion with the Lagrangian
\bea
\mathcal{L}_{a\gamma\gamma} \supset	\frac{1}{2}\partial_\mu a \, \partial^\mu a -\frac{1}{2}m^{2}_{a} a^{2} + \frac{g_{a\gamma\gamma}}{4} a F_{\mu\nu}\tilde{F}^{\mu\nu}\,
\eea
Here $F_{\mu\nu}$ is the field strength tensor of electromagnetic fields, and $\tilde{F}^{\mu\nu}$ is the dual field strength tensor. The axion coupling to the electromagnetic fields is $g_{a\gamma\gamma}$. In this study, we consider a generalized scenario where an axion-like particle is described with its mass $m_a$ and its coupling $g_{a\gamma\gamma}$ with the two being independent parameters. 
We focus on the axion-photon coupling to investigate its impact on stellar evolution through axion production, which contributes to stellar energy loss. A more detailed summary of the axion parameter space and the current constraints on $g_{a\gamma\gamma}$ can be found in~\cite{AxionLimits}.

The electromagnetic coupling of the axion induces its production channel via the Primakoff process $\gamma + \mathcal{Z}e \to a + \mathcal{Z}e$, where $\mathcal{Z}e$ represents target particles with charge $\mathcal{Z}$ in units of the electric charge $e$. These include ions (or nuclei) and electrons in the stellar medium. Photons from the hot stellar core are converted into axions in the presence of background electromagnetic fields provided by charged particles. However, the conversion rate is also influenced by the screening effect of the charge distribution in the dense plasma. As discussed in Sec.~\ref{sec: ClassicalCepheid}, energy loss in the helium core and hydrogen burning shell due to axion production could play a significant role in the evolution of Cepheids after helium ignition. This necessitates an accurate modeling of the axion Primakoff process around the burning shell. In particular, the high core temperature  and density inside the shell during the post-main-sequence phase lead to an increased plasma frequency and enhanced charge screening effect, both of which are critical for determining whether the stellar evolution curve crosses the instability strip and undergoes a complete blue loop. Therefore, the calculation of photon-axion conversion requires refined treatment by accounting for the influence of plasma frequency and electron degeneracy in the stellar medium. We follow the method developed in~\cite{Raffelt:1987yu} and refer readers to it for more details on the structure factor and screening scales. This method was recently employed in~\cite{Dolan:2022kul} to constrain the axion effect by analyzing the reduction in the $R$ parameter~\cite{1968Natur.220..143I} and the $R_2$ parameter~\cite{1989ApJ...340..241C} from globular cluster observations. In this approach, contributions from electrons and ions are treated independently, with each serving as a static, uniform charge density background for the other. We assume the axion mass is smaller than the temperature, making its production rate proportional only to the coupling. We also assume the axions can escape the star after being produced since the interaction strength is feeble, thus all axions contribute to additional stellar energy loss $\epsilon_a$. In terms of Cepheids, the main axion production is from the non-degenerate medium $\epsilon_{\rm nd}$ and degenerate medium $\epsilon_{\rm d}$. The non-degenerate medium contribution dominates in the outer stellar cells while the degenerate medium contribution becomes comparable in the inner core. We can model the total axion emission rate per unit mass as,
\begin{eqnarray} \label{axionenergylossrate}
    \epsilon_a = (1-w) \, \epsilon_{\rm nd} + w \, \epsilon_{\rm d},
\end{eqnarray}
where $w$ is a variable for the transition between the non-degenerate and the degenerate regimes. We choose $w$ to have the same functional form as in~\cite{Raffelt:1987yu},
\begin{equation}
    w=\frac{1}{\pi}\arctan(\zeta-3) + \frac{1}{2}.
\end{equation}
Here $\zeta$ is determined by the electron number density $n_e$ and temperature $T$ in each specific cell of \texttt{MESA} simulation (note that the local stellar cell temperature $T$ used in this section is different from the effective stellar temperature $T_{\rm eff}$ on the HR diagram),
\begin{equation}
    \zeta=\frac{(3 \, \pi^2 \, n_e)^{\frac{2}{3}}}{2 \, m_e \, T}.
\label{eq:zetaelectron}
\end{equation}
In the high density core with non-relativistic electron gas, the value of $\zeta$ approaches the Fermi energy $E_{\rm F}$ normalized by the temperature. The transition from the non-degenerate regime to the degenerate regime occurs when $\zeta$ increases to exceed $\mathcal{O}(1)$, indicating a considerable Fermi energy for degenerate electrons. We will address the axion production rate in different stellar components due to their corresponding screening scales in the following.

In the non-degenerate medium, all charged particles behave as sources of static Coulomb field, whose screening effect reduce the axion production rate. The axion production rate from non-degenerate medium is then
\begin{eqnarray}
    \epsilon_{\rm nd}=\frac{g^2_{a\gamma\gamma} \, T^7}{16\pi^2 \, \rho} \, y_1^2 \, F(y_0,y_1).
\label{eq:epsilonnd}
\end{eqnarray}
The density $\rho$ accounts for all stellar components. The function $F(y_0,y_1)$ takes the form
\begin{eqnarray}
    F(y_0,y_1)= \frac{1}{4\pi} \int_{y_0}^{\infty} {\rm d}y \, \frac{y^2\sqrt{y^2-y_0^2}}{e^{y}-1} \, I(y,y_0,y_1),
\label{eq:Ffunction}
\end{eqnarray}
where the dummy variable $y=\omega_\gamma/T$ accounts for the initial state photon energy, and
\begin{eqnarray}
    I(y,y_0,y_1)=\int_{-1}^{+1}{\rm d}x \, \frac{1-x^2}{(r-x)(r+s-x)}.
\end{eqnarray}
The variables $r$ and $s$ are chosen as
\begin{eqnarray}
    r&=&\frac{2y^2-y_0^2}{2y\sqrt{y^2-y_0^2}},\\
    s&=&\frac{y_1^2}{2y\sqrt{y^2-y_0^2}}.
\end{eqnarray}
The $y_0$ parameter represents the dependence on the plasma frequency
\begin{eqnarray}
    y_0=\frac{\omega_{\rm pl}}{T}.
\end{eqnarray}
The medium plasma frequency is 
\begin{eqnarray}
    \omega_{\rm pl}&=&\sqrt{\frac{4\pi\alpha n_e}{E_{\rm F}}},
\end{eqnarray}
and its numerical value can be approximated as~\cite{Raffelt:1996wa}
\begin{eqnarray}
    \omega_{\rm pl} \simeq 28.7~{\rm eV} \, \frac{(Y_e \, \frac{\rho}{{\rm g}\,{\rm cm}^{-3}})^{\frac{1}{2}}}{[1+(1.019\times10^{-6} \, Y_e \, \frac{\rho}{{\rm g}\,{\rm cm}^{-3}})^{\frac{2}{3}}]^{\frac{1}{4}}}.
\end{eqnarray}
Here $\alpha$ is the fine structure constant, and $Y_e$ is the number of electrons per nucleon. While the approximation $y_0\to 0$ is sometimes taken for horizontal branch stars, the effect of $y_0$ cannot be neglected for giant stars, where the plasma frequency is non-negligible compared to the temperature in the core regions. Therefore, we follow~\cite{Raffelt:1987yu} and keep the non-vanishing $y_0$ in our simulation. When Cepheids enter the blue loop, the helium core expands, and both its core temperature and plasma frequency decrease over time. In benchmark simulations, we observe that the value of $y_0$ decreases as the star crosses the instability strip during the blue loop. The $y_1$ parameter represents the dependence on the inverse Debye-H\"{u}ckel radius
\begin{eqnarray}
    y_1=\frac{k_{\rm nd}}{T}.
\end{eqnarray}
The Debye-H\"{u}ckel wavenumber $k_{\rm nd}$ of the non-degenerate medium accounts for all charged particle species $i$
\begin{eqnarray}
    k_{\rm nd}=\sqrt{\frac{4\pi\alpha}{T}\, \sum_{i}\mathcal{Z}_i^2 \, n_i},
\end{eqnarray}
with their number density $n_i$ and electric charge $\mathcal{Z}_i$. The $k_{\rm nd}$ and the momentum transfer in the Primakoff process determines the charge screening effect in the non-degenerate stellar medium. 

The degenerate medium produces axions with contributions from both electrons and ions
\begin{eqnarray}
    \epsilon_{\rm d} = \epsilon_{\rm ions} + \epsilon_{\rm e}.
\end{eqnarray}
In the limit of high electron degeneracy, the screening of ion charges from electron is negligible. Thus, the axion production from ions is calculated similarly to Eq.~\ref{eq:epsilonnd}, but with $y_1$ replaced by $y_2$ to account for a reduced screening level,
\begin{eqnarray}
    \epsilon_{\rm ions}=\frac{g_{a\gamma\gamma}^2 T^7}{16\pi^2 \rho} y_2^2 F(y_0,y_2).
\end{eqnarray}
The $y_2$ parameter contains only contributions from ions,
\begin{eqnarray}
    y_2=\frac{k_{\rm ions}}{T},
\end{eqnarray}
and the ionic Debye-H\"{u}ckel wavenumber is
\begin{eqnarray}
    k_{\rm ions}=\sqrt{\frac{4\pi\alpha}{T}\sum_{j}\mathcal{Z}_j^2 n_j}
\end{eqnarray}
the index $j$ runs for all ion species in the stellar medium. The degenerate electrons are correlated at a length scale set by the Thomas-Fermi wavenumber $k_{\rm TF}$, which is responsible for the charge screening in the core,
\begin{eqnarray}
    k_{\rm TF}=\sqrt{\frac{4\alpha m_e p_{\rm F}}{\pi}}
\end{eqnarray}
with $p_{\rm F}=(3\pi^2 n_{e})^{\frac{1}{3}}$ being the Fermi momentum of degenerate electrons. The degenerate electron correlation is much weaker than other particles, as can be seen with $k_{\rm TF} \ll k_{\rm ions} \lesssim k_{\rm nd}$. The axion production from degenerate electron is less influenced by the charge screening effect, but instead gets suppressed due to Pauli blocking
\begin{eqnarray}   \epsilon_{e}=\frac{g_{a\gamma\gamma}^2  T^7}{32\rho} \left(\frac{n_e}{3\pi\alpha m_e^3}\right)^{\frac{1}{2}} R_{\rm deg} \, y_3^3 \, F(y_0,y_3).
\end{eqnarray}
The reduction factor in the degenerate core is $R_{\rm deg}=1.5/\max{(1.5,\zeta)}$ with $\zeta$ defined in Eq.~\ref{eq:zetaelectron}. The parameter $y_3$ is determined with the Thomas-Fermi scale as
\begin{eqnarray}
    y_3=\frac{k_{\rm TF}}{T}.
\end{eqnarray}

\subsection{Energy Loss in \texttt{MESA} Simulation}

In this section, we discuss the implementation of axion processes in stellar evolution simulations with \texttt{MESA}. The axion effect has been incorporated into the \texttt{run\_star\_extras} module.
To implement the axion energy loss into the simulation, we express the function $F(y_0,y_1)$ in Eq.~\ref{eq:Ffunction} with a slowly varying function $G(y_0,y_1)$ as~\cite{Raffelt:1987yu,Dolan:2022kul}
\begin{eqnarray}
    F(y_0,y_1)=\frac{100}{1+y_1^2}\frac{1+y_0^2}{1+e^{y_0}}\,G(y_0,y_1)
    \label{eq:Gfunction}
\end{eqnarray}
We calculate the numerical values of $G(y_0,y_1)$ on a grid and tabulate the results in a data table, as shown in Table~\ref{tab:Gtable}.\footnote{We note that our evaluated values of the function $G(y_0, y_1)$ in Table~\ref{tab:Gtable} differ from those in Table II of~\cite{Raffelt:1987yu}, which use the same expressions. Our results are computed using the program \texttt{Mathematica 13.0.1.0}} The grid is selected to encompass a broad range of $y$ values for the stellar profiles. In a simulation of a $6 M_\odot$ star ($\alpha_{\rm ov}=0.1$ and $Z=0.014$) without axion emission, we observe the range: $-3.25 \lesssim \log y_0 \lesssim -0.95$, $-0.50\lesssim \log y_1 \lesssim 1.07$, $-0.64\lesssim \log y_2 \lesssim 0.93$, and $0.05\lesssim \log y_3 \lesssim 2.77$. The corresponding ranges of $y$ variables in a $8 M_\odot$ model is $-3.30 \lesssim \log y_0 \lesssim -1.09$, $-1.02\lesssim \log y_1 \lesssim 1.00$, $-1.98\lesssim \log y_2 \lesssim 0.86$, and $-0.02\lesssim \log y_3 \lesssim 2.76$.\footnote{Although the minimal value of $y_0$ falls outside the range presented in Table~\ref{tab:Gtable}, we have verified that the stellar evolutionary track  simulated with Table~\ref{tab:Gtable} remains consistent with simulations with an extended table, showing a difference of less than $0.07\%$ in the final stellar ages.} The numerical table is interpolated in our \texttt{MESA} simulations in order to calculate the axion emission rate $\epsilon_a$ in each cell.

\begin{table}[htbp]   
        \centering
        \label{tab:Gtable}
        \begin{tabular}{|c||c|c|c|c|c|c|c|c|c|c|}
        \hline
        $\log y$ &-3.0&-2.5&-2.0&-1.5&-1.0&-0.5&0.0&0.5&1.0&1.5  \\ \hhline{|=||=|=|=|=|=|=|=|=|=|=|}
        -2.0&0.2240&0.2481&0.2489&0.2513&0.2543&0.2182&0.0999&0.0227&0.0071&0.0032 \\
        \hline
         -1.5&0.1796&0.2010&0.2013&0.2033&0.2079&0.1981&0.0995&0.0227&0.0072&0.0032\\
         \hline
         -1.0&0.1375&0.1543&0.1548&0.1564&0.1603&0.1609&0.0961&0.0229&0.0072&0.0032\\
         \hline
        -0.5&0.1026&0.1162&0.1166&0.1178&0.1208&0.1233&0.0868&0.0245&0.0079&0.0035 \\
        \hline
         0.0&0.1054&0.1216&0.1220&0.1232&0.1264&0.1297&0.1000&0.0.039&0.0141&0.0063 \\
         \hline
         0.5&0.2253&0.2707&0.2716&0.2743&0.2815&0.2893&0.2317&0.1250&0.0703&0.0348  \\
         \hline
         1.0&0.4466&0.5686&0.5705&0.5763&0.5913&0.6085&0.4948&0.3206&0.3614&0.2919  \\
         \hline
         1.5&0.5463&0.7410&0.7436&0.7511&0.7707&0.7933&0.6485&0.4467&0.7586&1.5992  \\
         \hline
         2.0&0.5670&0.7740&0.7768&0.7846&0.8052&0.8289&0.6781&0.4721&0.8774&3.0475  \\
         \hline
        2.5&0.5674&0.7778&0.7806&0.7884&0.8091&0.8329&0.6815&0.4750&0.8923&3.3722   \\
        \hline
        3.0&0.5676&0.7781&0.7810&0.7888&0.8094&0.8333&0.6818&0.4753&0.8986&3.4090  \\
        \hline
        \end{tabular}
        \caption{Numerical values of the function $G(y_0, y_1)$ in Eq.~\ref{eq:Gfunction}. Each column corresponds to a different $\log y_0$, and each row corresponds to a different $\log y_1$.}
\end{table}

\section{Axions, Blue Loops, and Cepheid Variables} \label{sec: blueloopaxion}

In this Section, we present our main results on the incorporation of axion emission into the evolution of intermediate mass stars, particularly the blue loop stage. We begin with a general discussion of classical Cepheid variables, emphasizing the various factors that affect the morphology of the blue loop. We then introduce the axion model, and discuss how the blue loop changes when axion emission is incorporated.

\subsection{The Blue Loop}

Since the properties of the blue loop are crucial for understanding the ML and PL relations of Cepheids, we summarize the general features of this evolutionary stage. Our discussion applies to a typical intermediate mass star, although we will show detailed results for a $6 M_\odot$ benchmark.

During the star's evolution on the main sequence, hydrogen burning releases nuclear binding energy by its conversion to helium. The rate of energy production $\epsilon_H$ depends both on the local temperature as well as the hydrogen abundance. The timescale of central hydrogen burning scales inversely with the stellar mass as $\tau_H \sim M^{-2.5}$, for an approximate ML relation $L \sim M^{3.5}$. Moreover, the temperature sensitivity of the CNO cycle implies that in more massive stars helium production is more concentrated toward the center. While the temperature increases in the center, the hydrogen abundance decreases. A useful parameter to study the spatial variation of various physical quantities is the ratio of the enclosed to the total stellar mass $m_{\rm enc}$. With increasing stellar mass, the hydrogen profile peaks at larger values of $m_{\rm enc}$ since the mass of the helium core increases. Convective overshoot and semiconvection render the above simplified description more complex, necessitating numerical simulations; however, the broad features persist.

At the onset of the post-main sequence stage, an approximately isothermal helium core exists, surrounded by a hydrogen envelope with a hydrogen burning shell at its lowest layers. The shell then induces the so-called ``mirror principle'' of radial motion of material inside stars, wherein the core and envelope evolve oppositely in extent on either side of the burning shell: an expansion of the envelope above the shell source is accompanied by a contraction of the layers below.  The envelope expansion rapidly converts the star into a red giant, placing it on the RGB near the Hayashi line. The contraction of the core leads to heating; when the temperature reaches $\sim 10^8$ K, the helium ignites and the core stops contracting.

The subsequent evolution of the star, in particular whether it embarks on a blue loop or not, is primarily determined by the core potential, defined in Eq.~\ref{corepot}. Numerical simulations for various profiles indicate the following parametric behavior of $h = h(\Delta X, \Delta m)$ \cite{Kippenhahn:2012qhp}
\be
h \, = \, e^{{\rm const.} \cdot \Delta X \,\Delta m}\,\,.
\ee
The function $h$ can be influenced by various factors, such as the mixing length in the superadiabatic layer which determines the mixing of the  outer convection zone. Under the assumption of full mixing,  the hydrogen profile can exhibit a plateau, with a step-function like drop in the hydrogen abundance at the location of the shell. In this scenario, one can expect 
\be
\Delta m \rightarrow 0, \,\,\, {\rm and} \,\,\, h \rightarrow 1\,\,.
\ee

The star's initial transit to the RGB involves an expansion of the envelope and contraction of the core, due to the mirror principle mentioned before. Due to the decrease in core radius $R_c$, the core potential $\phi_c$ increases during this stage, following Eq.~\ref{corepot}. As long as the core potential remains above the critical value $\phi_{\rm crit}$ the star remains on the RGB and ascends it. The value of $\phi_{\rm crit}$ can be determined by simulations; for example,  \cite{Kippenhahn:2012qhp} provides values of  $\log{(\phi_{\rm crit})} = 0.83, 0.93,$ and $0.99$ for stellar mass $M = 3 M_\odot, 5 M_\odot,$  and $7 M_\odot$, respectively. 

\begin{figure}[b]
  \centering
    \includegraphics[width=0.47\textwidth]{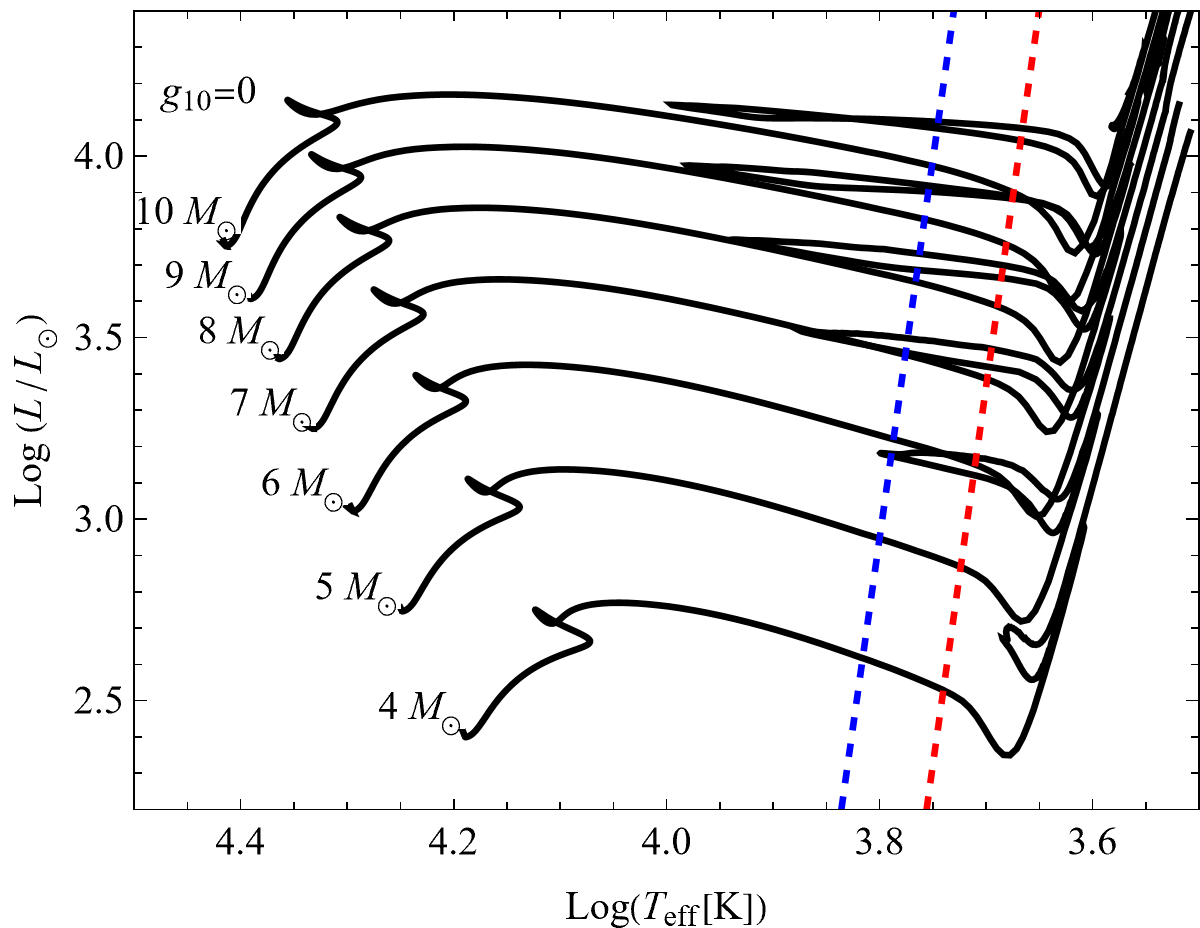}\qquad
    \includegraphics[width=0.47\textwidth]{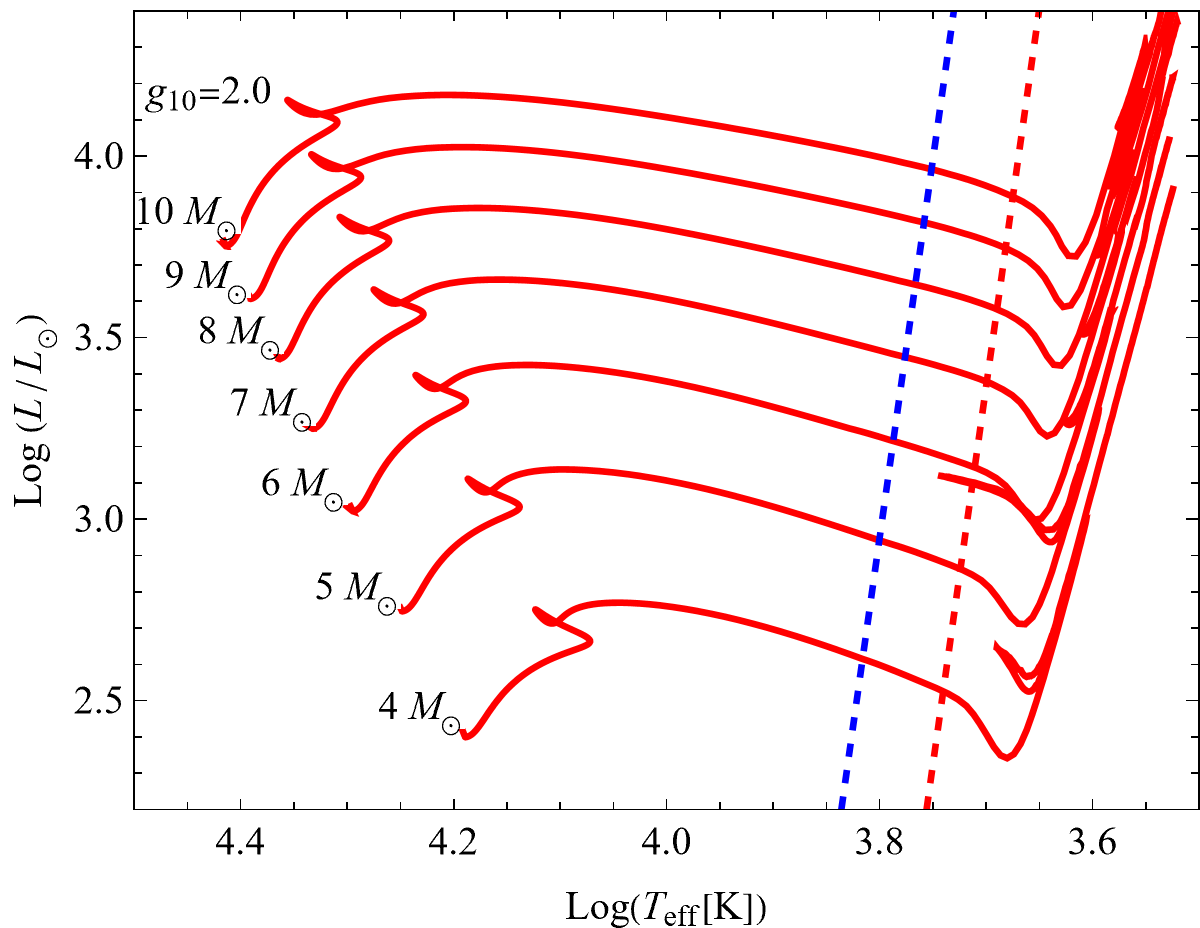}
  \caption{Evolutionary tracks for two different axion couplings $g_{10} = 0$ (left panel) and $g_{10} = 2$ (right panel) for a mass range of $4 - 10$ $M_{\odot}$. The metallicity is set to $Z = 0.014$ and the overshoot to $\alpha_{\rm ov} = 0.1$. }
  \label{fig:varevotracks}
\end{figure}

We display typical evolutionary tracks for $4-10 M_\odot$ stars on the left panel of Fig.~\ref{fig:varevotracks}. We take $\alpha_{\rm ov} =0.1$ above the hydrogen core. The models in Fig.~\ref{fig:varevotracks} have no rotation and are terminated midway through the AGB stage. The black tracks on the left panel correspond to the case where the axion-photon coupling 
\begin{equation}
g_{10}\equiv \frac{g_{a\gamma\gamma}}{10^{-10} \,{\rm GeV}^{-1}} 
\end{equation}
is set to zero. It is clear that when the coupling is turned off all models except the 4 $M_{\odot}$ case reach the instability strip, and more massive stars have more prominent blue loops. As noted above, this is primarily due to the fact that more massive stars have larger values of $\phi_{\rm crit}$. We note that the blue loop is inhibited for the $10 M_\odot$ (and heavier) models, due to the limited inward penetration of the convective envelope 
- a point further discussed in Section~\ref{overshootuncert}  \cite{10.1093/mnras/stu2666}.

As $g_{10}$ is switched on and increased from 0 to 2, blue loops in the more massive models disappear first while the models with lower mass,  e.g. 4 and 5 $M_{\odot}$, persist. We discuss this point in more detail in Sec.~\ref{blueloopmass}. The right panel of Fig.~\ref{fig:varevotracks} displays the evolutionary tracks with the axion-photon coupling set to $g_{10} = 2$. Other major periods of stellar evolution, for example the main sequence, RGB, and AGB stages, appear more or less unchanged when the axion-photon coupling is increased. We note that there will in fact be a 
noticeable effect in the core of the star in the post blue loop evolution, as shown in Appendix~\ref{postblueloop}, where we plot the core temperature versus core density with and without axions. However, these differences in the core properties are not reflected in the overall evolutionary track of the star in the AGB, reinforcing the fact that the blue loop stage is the most sensitive stage of the evolution.

\begin{figure}[b]
  \centering
    \includegraphics[width=0.62\textwidth]{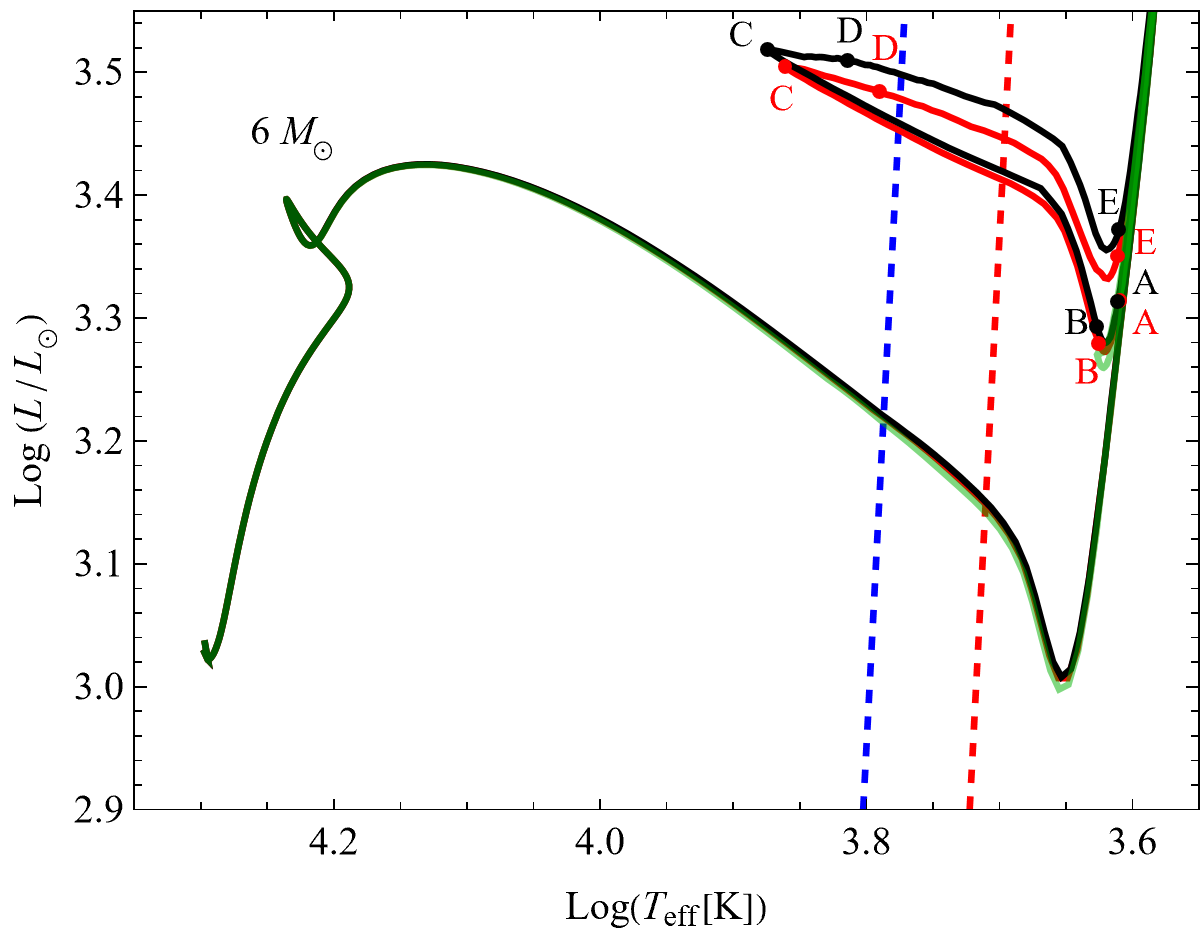}
   \caption{
   Evolutionary tracks of a 6 $M_\odot$ model for $g_{10} = 0 $ (black), $g_{10} = 1 $ (red), and $g_{10} = 2 $ (green). ``A'' denotes the beginning of the blue loop (where ${\rm Log}\, T_{\rm eff} \sim 3.605$), ``C'' the bluest point with the maximum effective temperature,  ``E'' the end, while the locations of ``B'' and ``D'' are determined by taking the average age of the star between ``A''  and ``C'', and ``C'' and ``E'', respectively. The metallicity is set to $Z = 0.014$ and the overshoot to $\alpha_{\rm ov} = 0.1$. }
   \label{fig:6MLT}
\end{figure}

The trajectory of a $6 M_\odot$ star is displayed separately in Fig.~\ref{fig:6MLT}. The black curve corresponds to the case where the axion-photon coupling is set to zero. To better display the evolution of the star during the loop, we label the start of the blue loop stage by ``A'' (where $\log T_{\rm eff} \sim 3.605$), and the bluest point with the maximum effective temperature by ``C''. ``E'' denotes the end of the blue loop, while the locations of ``B'' and ``D'' are determined by taking the average age of the star between ``A''  and ``C'', and ``C'' and ``E'', respectively. It should be noted, therefore, that ``B'' and  ``D'' indicate the midpoints of age, not the effective temperature. It is evident that while the phase from ``A''  to ``B'' produces a relatively small increase in the effective temperature, the same amount of time from ``B''  to ``C'' produces a remarkable increase in the effective temperature. A similar phenomenon occurs on the reverse passage. The red and green curves correspond to axion-photon couplings set to $g_{10} = 1$ and $g_{10} = 2$, respectively. It is evident that while $g_{10} = 1$ shortens the blue loop noticeably, the loop is entirely eliminated for $g_{10} = 2$.

\subsection{The Mirror Principle}

Given a star with an initial value of the core potential $\phi_{c, {\rm init}}$, this value can fall below  $\phi_{\rm crit}$ due to a reduction in $h = h(\Delta X, \Delta m)$ as the shell burns, or an  increase in the core radius, or both. This initiates the leftward movement along the blue loop. During this stage, the core expands, leading to a further drop in the core potential. Due to the mirror principle, the envelope contracts, moving the star further to the left on the HR diagram. At some point, the core potential reaches its minimal value $\phi_{c, {\rm min}}$. The core subsequently contracts and the envelope expands, leading to an increase in $\phi_{c}$. The star moves to the right in the HR diagram, completing the blue loop. After  $\phi_{c}$ becomes larger than $\phi_{\rm crit}$, the star exits the blue loop and ascends the AGB.

\begin{figure}[h]
  \centering
    \includegraphics[width=0.6\textwidth]{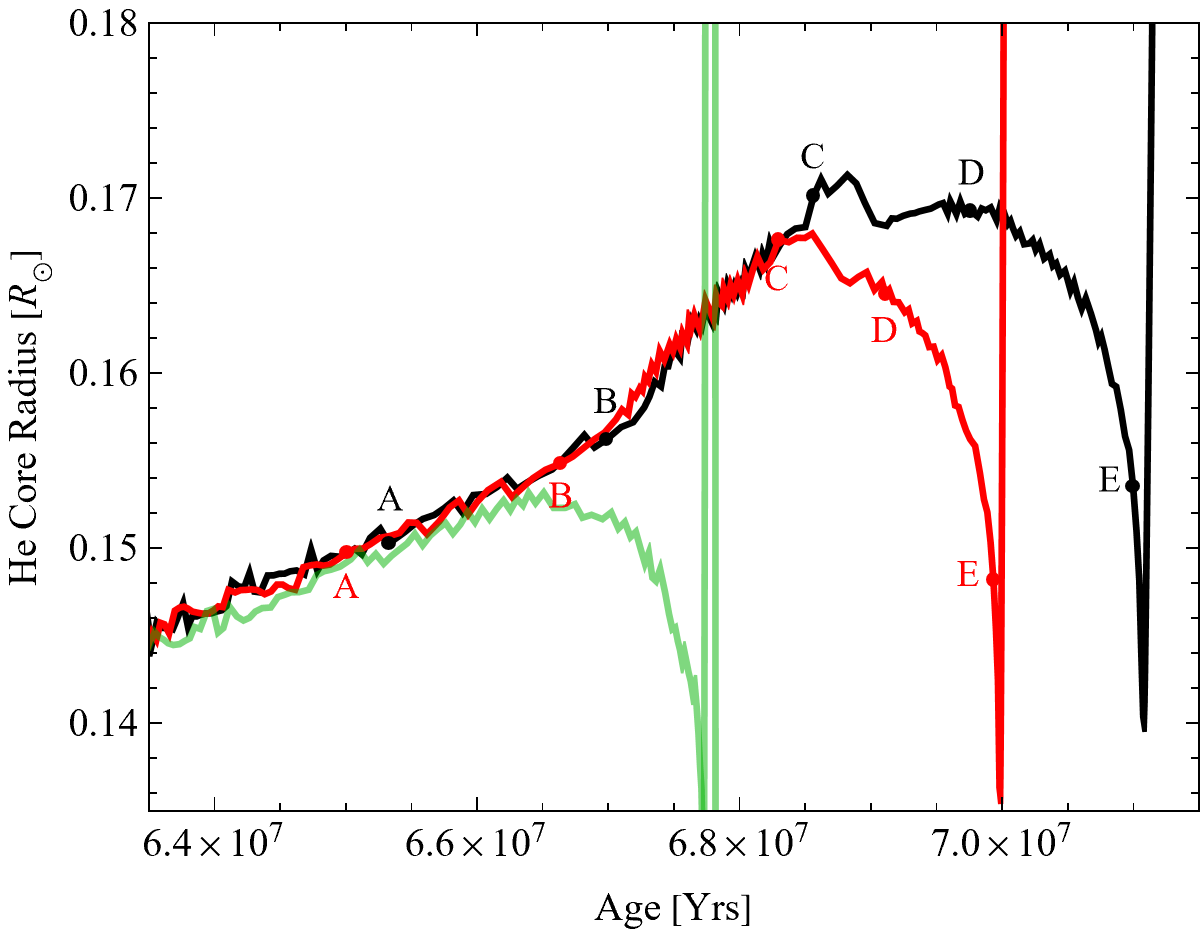}\\
    \includegraphics[width=0.6\textwidth]{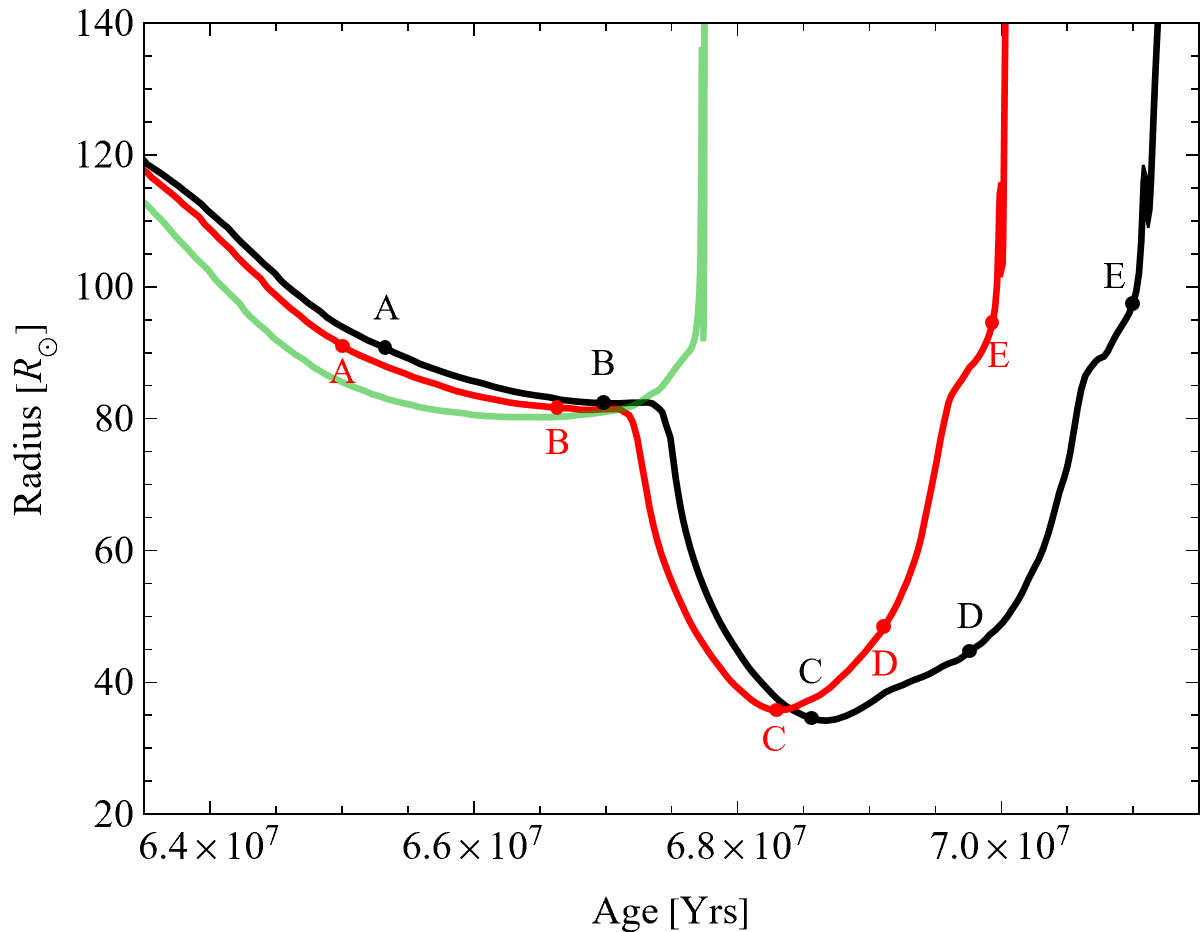}\\
  \caption{Top panel: The helium core radius as a function of the stellar age for a 6 $M_\odot$ model. The axion couplings are $g_{10} = 0 $ (black), $g_{10} = 1 $ (red), and $g_{10} = 2 $ (green). ``A'' denotes the beginning of the blue loop (where $\log T_{\rm eff} \sim 3.605$), ``C'' the bluest point with the maximum effective temperature,  ``E'' the end, while the locations of ``B'' and ``D'' are determined by taking the average age of the star between ``A''  and ``C'', and ``C'' and ``E'', respectively.  Bottom panel: The stellar radius as a function of the stellar age for the same model. For both panels, the metallicity is set to $Z = 0.014$ and the overshoot to $\alpha_{\rm ov} = 0.1$.}
  \label{fig:Radius}
\end{figure}

The black curves in Fig.~\ref{fig:Radius} show the evolution of the core (top panel) and envelope (bottom panel) during the blue loop phase, in unit of the solar radius $R_\odot$, for a $6 M_\odot$ star. The age in all figures signifies time elapsed since the zero age main sequence stage. From the top panel, it is evident that at the onset of the blue loop (``A'' around age $\sim 6.5 \times 10^7$ yrs) the core radius starts increasing. The radius reaches a maximum value of $\sim 0.17 R_\odot$ at the bluest point (``C'' at age $\sim 6.9 \times 10^7$ yrs) after which it rapidly shrinks until it reaches a radius of $\sim 0.15 R_\odot$ (``E'' at age $\sim 7.2 \times 10^7$ yrs). From the bottom panel, it is evident that the envelope follows the mirror principle at each stage of the progression. It should be noted that the vertical axis of the top panel, depicting the core radius, is substantially magnified (ranging from $0.135 - 0.18 R_\odot$) compared to the vertical axis of the bottom panel, depicting the extent of the envelope (ranging from $20 - 140 R_\odot$). 

The red and green curves in Fig.~\ref{fig:Radius} correspond to the cases where the axion-photon coupling is set to $g_{10} = 1$ and $g_{10} = 2$, respectively. From the top panel, it is evident that with increasing $g_{10}$ the maximum value of the core radius is reduced compared to the case when axions are absent. Indeed, for the case of $g_{10} = 1$, the  value of the core radius at the position ``C'' is $R_{c} \sim 0.166 R_\odot$, compared to the value at the position of the black triangle $R_{c} \sim 0.17 R_\odot$, a reduction of $\sim$ 2\%. The value of the core radius is even further reduced in the case of $g_{10} = 2$, with $R_c \sim 0.150$, a reduction of $\sim 11\%$ compared to the black peak. Increasing the axion-photon coupling leads to an increase in energy loss which causes the star to go through helium burning more rapidly and reduces the size of the core compared to the case when $g_{10} = 0$. Since the energy loss $\epsilon_a \propto g_{10}^{2}$, increasing the coupling from $g_{10}=1$ to $g_{10} =2 $ correspond to a four-fold increase in energy loss. From the red and green curves in the bottom panel, it is evident that the envelope follows the mirror principle for each axion-photon coupling.

\subsection{The Hydrogen Profile and the Core Potential}

\begin{figure}[h]
  \centering
    \includegraphics[scale=0.91]{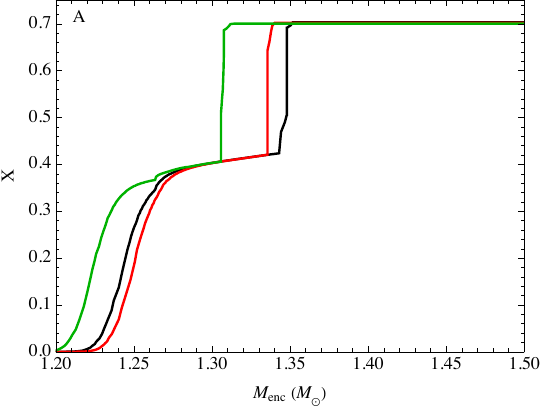}
    \includegraphics[scale=.91]{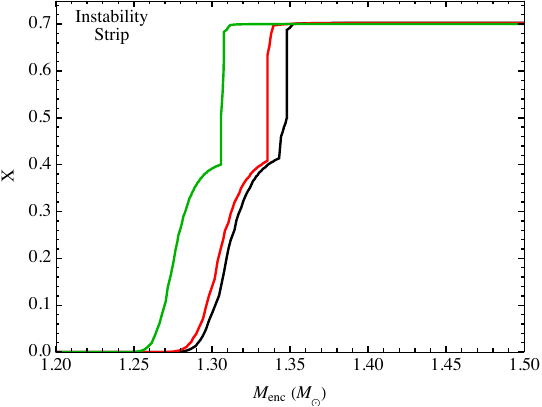}
    \includegraphics[scale=.91]{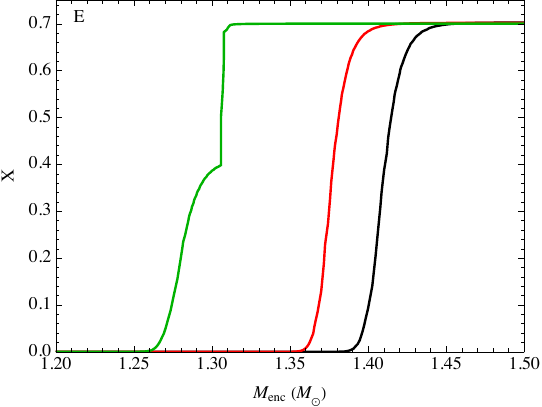}
  \caption{The hydrogen mass fraction $X$  profile of a 6 $M_\odot$ model before (top panel), during (middle panel) and after (bottom panel) the blue loop stage, for axion couplings $g_{10} = 0 $ (black), $g_{10} = 1 $ (red), and $g_{10} = 2 $ (green). The metallicity is set to $Z = 0.014$ and the overshoot to $\alpha_{\rm ov} = 0.1$.}
  \label{fig:hydprofile}
\end{figure}

The  hydrogen profile as a function of the enclosed mass, $m_{\rm enc}$, is displayed in Fig.~\ref{fig:hydprofile}, for three stages of the blue loop: stage ``A'' (top panel), during the transit through the instability strip (middle panel), and stage ``E'' (bottom panel). The black, red, and green curves indicate the cases of $g_{10} = 0 $, $g_{10} = 1 $, and $g_{10} = 2 $, respectively. The black profile shows the expected behavior in the absence of axions. Reading from left to right in the top or middle panels, one has: $(i)$ the convective inner core where helium converts to carbon and the radiative outer core (these are flat lines on the hydrogen profile); $(ii)$ the sharp first increase in the hydrogen profile, which  marks the hydrogen-burning shell; $(iii)$ the second more steady increase where the old convective core retreats; and $(iv)$ finally, the jump which flat-lines into the envelope composition on the right.

In stage ``A'', a jump in the hydrogen profile  appears due to the first dredge-up, when the convective zone of the envelope penetrates into the region where hydrogen is being burnt in the core. The old composition profile is truncated and matter from the core is redistributed to the envelope. In the middle panel, both the inner core and the burning shell move outwards when plotted against the enclosed mass fraction. In the final panel, the burning shell attenuates the zone where the composition steadily changes, and reaches the jump. The presence of axions during the instability strip preserves the overall morphology of the hydrogen profile, shifting it leftward towards smaller enclosed mass. 

It is clear how the function $h$ would behave, based on Fig.~\ref{fig:hydprofile}. Following the black curve with $g_{10} = 0 $, it is clear that as the blue loop progresses, the value of $\Delta m$ decreases from $\Delta m \sim 0.1$ (top panel), $\Delta m \sim 0.06$ (middle panel) until the change in the hydrogen profile becomes almost smooth and $\Delta m \sim 0.01$ (bottom panel). Since $\Delta X \sim 0.7$ remains constant in the three stages, it is clear that the function $h$ decreases from its value at the onset of the blue loop, ultimately reaching $h \rightarrow 1$ at the end. It is, therefore, a combination of the fact the $h$ decreases and $R_c$ increases in Fig.~\ref{fig:Radius} that leads to a decrease in $\phi_c$ and a blue loop is initiated. From Fig.~\ref{fig:hydprofile}, it is clear that non-zero axion couplings do not produce a substantial difference in the values of $\Delta m$ and $\Delta X$, and hence in the behavior of $h$,  compared to the case where the couplings are zero. The main difference is that as the axion coupling becomes stronger, the same qualitative jump in the hydrogen profile occurs at smaller $m_{\rm enc}$, reflecting a smaller core size. We note that the core mass $M_c$ increases in both in the absence as well as in the presence of the axion coupling.

\begin{figure}[h]
  \centering
    \includegraphics[width=0.65\textwidth]{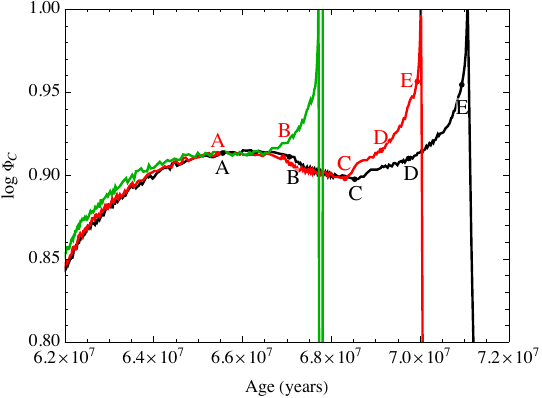}
   \caption{Core potential as a function of age for a 6 $M_\odot$ model with $g_{10} = 0 $ (black), $g_{10} = 1 $ (red), and $g_{10} = 2 $ (green). Different evolutionary stages are labeled with “A” through “E,” as depicted in Fig.~\ref{fig:6MLT}. The metallicity is set to $Z = 0.014$ and the overshoot to $\alpha_{\rm ov} = 0.1$. }
   \label{fig:corepot}
\end{figure}

Finally, the evolution of the core potential as a function of the age of the star is displayed in Fig.~\ref{fig:corepot}. The black curve displays the case where the axion coupling is zero. It is evident that at the beginning of the blue loop (``A'' at age  $\sim 6.5 \times 10^7$ yrs) the core potential is at its maximum value of $\sim 0.91$, which is larger than $\phi_{\rm crit}$ for a $6 M_\odot$ star. Thereafter, the core potential decreases, falls below $\phi_{\rm crit}$, reaches its minimum value and subsequently increases. On the other hand, the reduction in the core potential is less pronounced for the red curve, and almost completely absent for the green curve. This demonstrates that the blue loop is eliminated for $g_{10} = 2$ because  the core potential remains too high, which is primarily driven by the fact that the core radius $R_c$ remains too small.

\subsection{Blue Loop Elimination as a Function of Stellar Mass} \label{blueloopmass}

More massive stars lose their blue loops first compared to less massive stars, when axions are switched on, even though the massive ones have more robust loops to begin with (Fig.~\ref{fig:varevotracks}). Constraints on axions coming from stars of different masses have different levels of robustness and are of different strengths, and we explore these questions in this section.

\begin{figure}[t]
  \centering
    \includegraphics[width=0.65\textwidth]{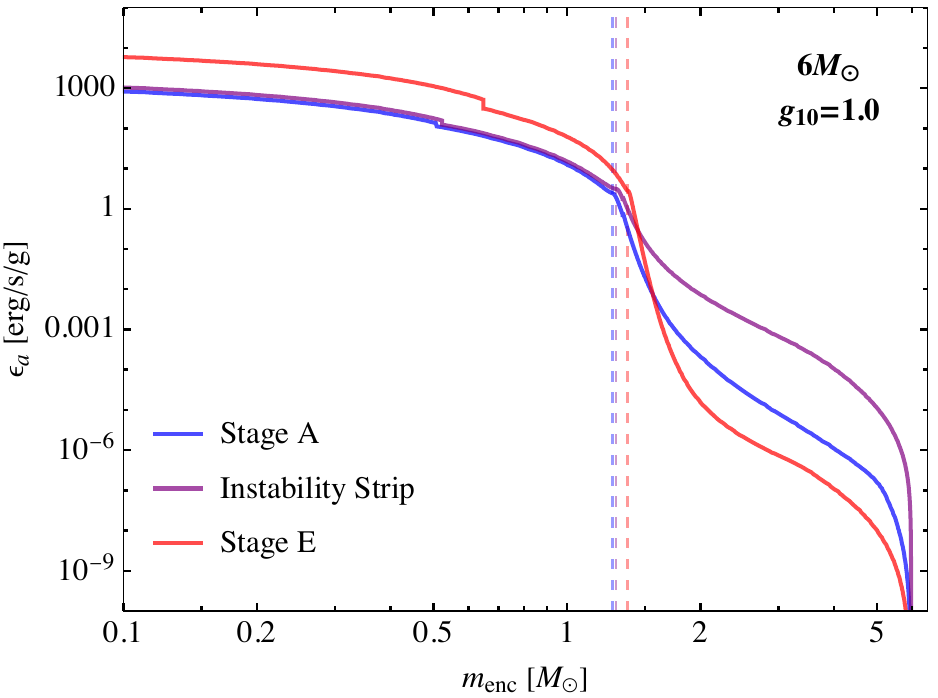}\\
      \includegraphics[width=0.65\textwidth]{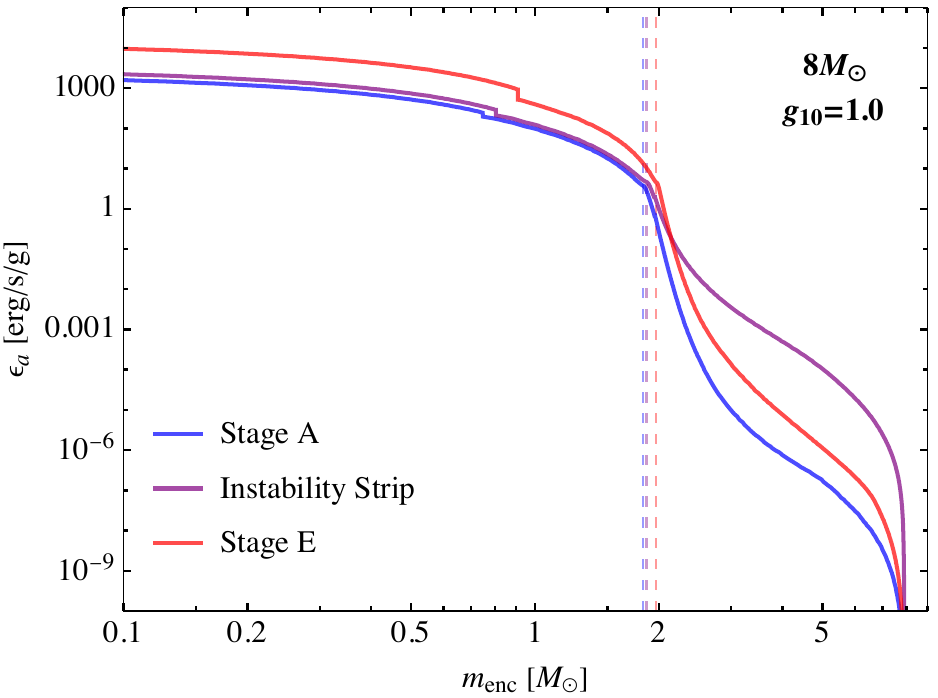}
   \caption{The axion energy loss rate from Eq.~\ref{axionenergylossrate}, as a function of the enclosed mass for a $6 M_\odot$ (top) and $8 M_\odot$ model (bottom), for Stage ``A'' (blue), during the transit through the instability strip (purple), and Stage ``E'' (red) of the blue loop stage. The axion coupling is $g_{10} = 1 $.  The metallicity is set to $Z = 0.014$ and the overshoot to $\alpha_{\rm ov} = 0.1$.}
   \label{fig:axionemission6and8}
\end{figure}

The emission rates within a stellar cell, plotted as a function of the enclosed mass, are shown in Fig.~\ref{fig:axionemission6and8} for a $6M_\odot$ model (top panel) and a $8M_\odot$ model (bottom panel). The instantaneous axion emission rate is calculated using Eq.~\ref{axionenergylossrate} and the stellar profile generated from the \texttt{MESA} simulation. We assume an axion coupling of $g_{10}=1.0$, a metallicity of $Z=0.014$, and an overshoot parameter of $\alpha_{\rm ov}=0.1$. The color curves represent emission rates at different stages of the star’s evolution: before (blue), during (purple), and after (red) the blue loop phase along the evolutionary track.  Several features are evident: $(i)$ Firstly, the helium burning core overwhelmingly dominates axion production.  Energy loss  within the  core (boundary indicated by the dashed line) dominates due to its high temperature and falls off dramatically by several orders of magnitude  beyond $m_{\rm enc} \sim 10\%$ of the star mass, for all stages before, during, and right after the blue loop. For larger values of the axion coupling, axion emission leads to excessive energy loss in the core, accelerates helium burning, and eventually eliminates the blue loop phase. $(ii)$ Secondly, axion emission is more efficient in heavier stars. We compare the local emission rate of a representative stellar cell, located where $m_{\rm enc}$ is $10\%$ of the star mass, between the $8 M_\odot$ model and the $6 M_\odot$ model shown in Fig.~\ref{fig:axionemission6and8}. The corresponding ratios are about $2.2$ before (blue), $2.9$ during (purple), and $2.0$ (red) after the blue loop phase. Additionally, the ratios of the total axion emission power between the $8 M_\odot$ model and the $6 M_\odot$ model are about $2.5$ (blue), $2.8$ (purple), and $2.1$ (red), respectively. The variation in emission rates calculated using Eq.~\ref{axionenergylossrate} arises from the core mass difference by a factor of approximately $1.4$, and from differences in core temperatures and densities. In our simulation, we find the ratios of the highest core temperatures are $1.07$ (blue), $1.09$ (purple), and $1.03$ (red). Therefore, we expect the sensitivity to $g_{10}$ to improve by a factor of a few as the stellar mass increases in our benchmarks.

\begin{figure}[b]
  \centering
    \includegraphics[width=0.75\textwidth]{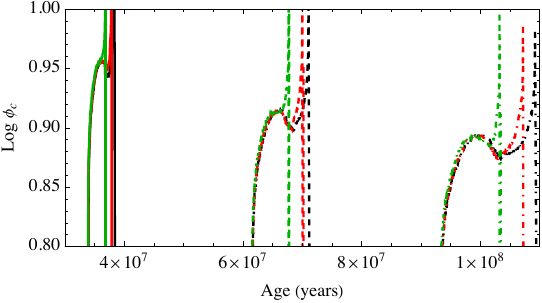}
   \caption{Core potential as a function of age for 5 $M_\odot$ (right), 6 $M_{\odot}$ (middle), and 8 $M_\odot$ (left) models with $g_{10} = 0 $ (black), $g_{10} = 1 $ (red), and $g_{10} = 2 $ (green). The metallicity is set to $Z = 0.014$ and the overshoot to $\alpha_{\rm ov} = 0.1$.}
   \label{fig:corepotmass}
\end{figure}

In Fig.~\ref{fig:corepotmass}, we show the evolution of the core potential as a function of the age of the star, for several selections of stellar masses: 5 $M_\odot$ (right, dashdotted), 6 $M_\odot$ (middle, dashed), and 8 $M_\odot$ (left, solid). The black, red, and green curves for a given mass correspond to $g_{10} = 0$, $g_{10} = 1$, and $g_{10} = 2$. In the absence of axions (black curves),   the value of $\phi_{\rm crit}$ increases with increasing stellar mass \cite{Kippenhahn:2012qhp}. For a blue loop to be initiated, $\phi_c$ must fall below $\phi_{\rm crit}$;  massive stars typically satisfy $\phi_{c,{\rm init}} \sim \phi_{\rm crit}$ on the RGB and a blue loop is initiated and relatively robust. On the other hand,  for less massive stars, $\phi_{\rm crit}$ is already so low that the beginning value of $\phi_c$ may satisfy $\phi_{c,{\rm init}} \gg \phi_{\rm crit}$ and the blue loop never gets initiated, or is heavily dependent on the composition, rotation, and convection. The situation is quite different once the axion coupling is turned on. Clearly, for $g_{10} = 2$ (green), the core potential in the case of the 5 $M_\odot$ star satisfies $\phi_c < \phi_{\rm crit}$, initiating the blue loop, while for heavier stars this condition is violated and the blue loop disappears.

\section{Axion Constraints from the Existence of Massive Cepheids}\label{sec: axionconstrmassive}

In this section, we establish constraints on the axion-photon coupling using the criterion of blue loop elimination in Cepheid models corresponding to both existing observations and future expectations. 

\subsection{Constraints from Observed Cepheids and Simulations}

\begin{figure}[b]
  \centering
    \includegraphics[width=0.55\textwidth]{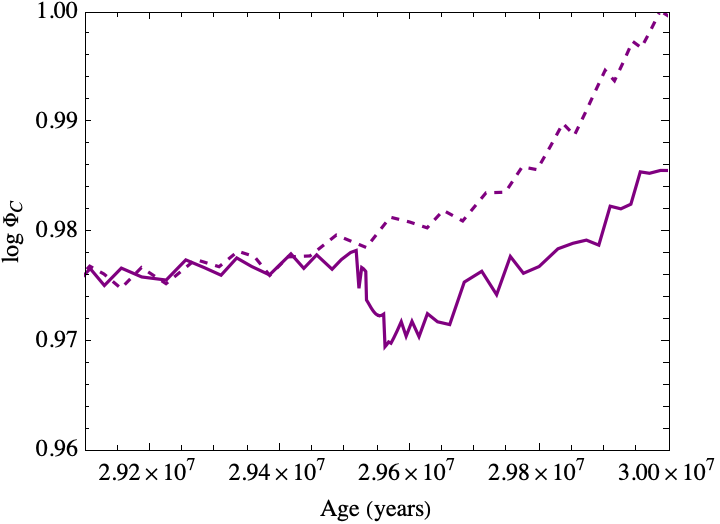}
  \caption{ Core potential of a 9 $M_\odot$ model for $g_{10} = 0.83 $ (solid) and $g_{10} = 0.835 $ (dashed). The metallicity is set to $Z = 0.014$ and the overshoot to $\alpha_{\rm ov} = 0.1$.}
  \label{fig:9msunconstraints}
\end{figure}

\begin{figure}[h!]
  \centering
     \includegraphics[width=0.45\textwidth]{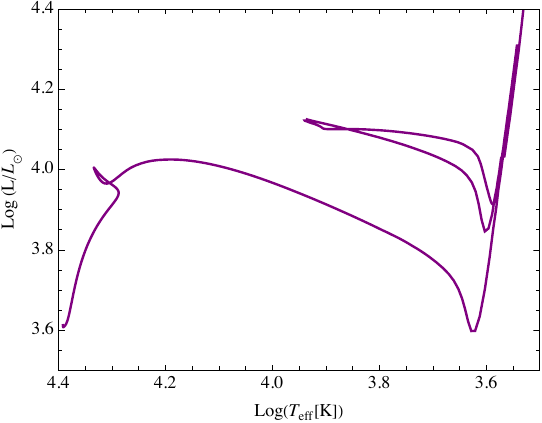}
        \includegraphics[width=0.45\textwidth]{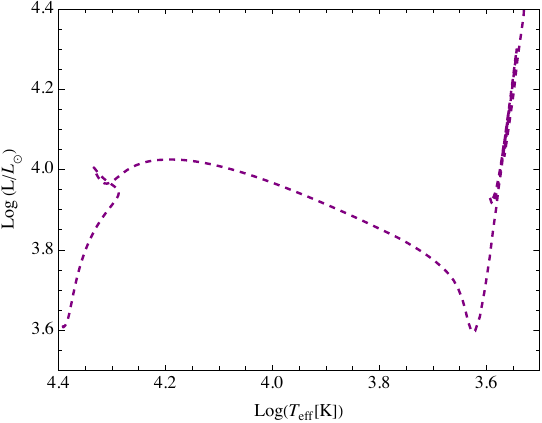}
  \caption{HR diagrams corresponding to the core potentials depicted in Fig.~\ref{fig:9msunconstraints}. The left panel corresponds to $g_{10}=0.83$ while the right panel corresponds to $g_{10}=0.835$. Both panels are for a 9 $M_{\odot}$ model. The metallicity is set to $Z = 0.014$ and the overshoot to $\alpha_{\rm ov} = 0.1$.}
  \label{fig:9msunHRdiags}
\end{figure}

It is evident from the discussion in Sec.~\ref{blueloopmass} that requiring the existence of  blue loops in more massive stars leads to more stringent constraints on $g_{10}$. Not only does the blue loop in more massive stars vanish for smaller values of $g_{10}$, the constraints obtained are more robust against assumptions in composition, rotation, and convection, since the blue loop without axions is more robust. In Fig.~\ref{fig:9msunconstraints}, we show the evolution of the core potential as a function of age for a $9 M_\odot$ model. The solid curve corresponds to  $g_{10} = 0.83 $, while the dashed curve corresponds to $g_{10} = 0.835$. It is clear that the blue loop disappears (stops before reaching the edge of the red instability strip) for $g_{10} = 0.835$, a fact that we have verified by the star's track on the HR diagram in Fig.~\ref{fig:9msunHRdiags}.

While stopping the evolution right at the edge of the red strip is our criterion, for heavier stars this criterion is sometimes difficult to fulfill since the evolution becomes highly sensitive to changes in energy loss. A small change in the coupling, at the level of $\Delta g_{10} \sim \mathcal{O}(10^{-4})$, can completely remove the blue loop instead of stopping it at the red edge. To address this, we create two models: one where the blue loop is entirely eliminated with a slightly higher coupling, and another where the blue loop persists with a lower coupling. For example, in the 8 $M_{\odot}$ model, the blue loop is eliminated at $g_{10}=1.2254$, while it remains at $g_{10}=1.2253$. Instead of selecting a single model to calculate the axion constraint presented later in Eq.~\ref{linift}, we employ both models. This approach is applied to models with masses greater than 7 $M_{\odot}$.

At this stage, the fact that values of $g_{10}$ must be so finely tuned to eliminate the blue loop might give reasons to pause; the elimination, at that level of precision, may be an artifact of the stellar evolution interpolation in \texttt{MESA}, rather than an accurate representation of the effect of axions. To a certain extent, this ambiguity is difficult to resolve without adopting a finer resolution for the cell thickness in the simulations, or upgrading the simulations in other non-trivial ways. The elimination of the blue loop for such extremely small changes of $g_{10}$ could also be due to underlying physical processes, which would require further study and more extensive simulations. However, it should be kept in mind that this is not a precision study: the blue loop does in fact vanish, for example in the case of the $9 M_\odot$ model, between $g_{10} = 0.83$ and $g_{10} = 0.84$, or, being even less precise, between $g_{10} = 0.8$ and $g_{10} = 0.9$. Presenting our results with fewer significant digits would not alter the overall physics statement that axion energy loss causes the blue loop to vanish, or improve/diminish the strengths of our constraints in a very meaningful way. As such, we choose to display as many significant digits in $g_{10}$ as we obtain from the current \texttt{MESA} simulation.

\begin{figure}[h]
  \centering
    \includegraphics[width=0.58\textwidth]{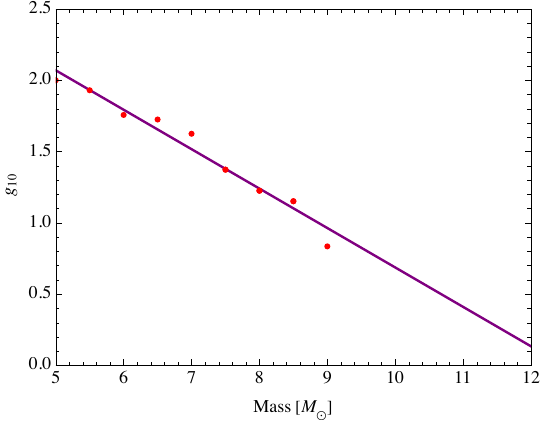}
  \caption{Values of the axion coupling $g_{10}$ at which the blue loop just disappears, for models simulated in the current work (red dots). The purple line provides a linear fit. }
  \label{fig:varyingmassconstraints}
\end{figure}

In Fig.~\ref{fig:varyingmassconstraints}, we plot the value of $g_{10}$ at which the blue loop just stops at the red edge of the instability strip, for models simulated in our work. The purple line provides a linear fit. We were unable to generate blue loops for models with masses larger than $9 M_\odot$ on \texttt{MESA} for $\alpha_{\rm ov} = 0.1$, even with $g_{10} = 0$. The inability of stellar evolutionary codes to produce blue loops for Cepheids above $9 M_\odot$ as long as overshoot is incorporated, even with reasonable conditions for other input parameters, has been observed generally across various simulation suites - for more details, we refer to \cite{10.1093/mnras/stu2666}. The blue loop is inhibited for $10 M_\odot$ and heavier  models at solar metallicities due to the limited inward penetration of the convective envelope.

At this point, one can begin to translate the results we have obtained into statements about constraints on the plane of $(g_{a\gamma\gamma}, m_a)$. The constraints come from requiring that a tentative Cepheid loses its blue loop due to axion emission; the criterion hinges on the behavior of the core potential, as displayed in Fig.~\ref{fig:9msunconstraints} for the case of a $9 M_\odot$ model. The requirement is that a blue loop should be curtailed just as it reaches the outer edge of the red edge of the instability strip. One can take these results, and compare them against Cepheids that have \textit{actually} been observed, preferably with masses determined by methods that are independent of any processes involving axions. In Table~\ref{tab:cepheidsobs}, we summarize the current observational status of galactic Cepheids with dynamical mass determination. We note that Large Magellanic Cloud candidates, for example OGLE-LMC-CEP0227 \cite{2010Natur.468..542P} with mass $3.74 M_\odot$ are not considered in our work. 

\begin{table}[h]
\centering
\caption{Observed Cepheids with Determined Masses}
\label{tab:cepheidsobs}
\vspace{3pt}
 \begin{tabular}{|c | c | c | c | c | } 
 \hline\hline
 Name & Mass ($M_\odot$) & Period (days)  & Mass Determination & Location \\ \hline
 
 SU Cyg   & 5.9 \cite{1986ESASP.263..405R} & 3.8455 \cite{Wenger:2000sw} & Dynamical & galactic \\ \hline

  V350 Sgr   & 5.2 \cite{Evans_2018} & 5.154 \cite{Wenger:2000sw} & Dynamical & galactic \\ \hline

  S Mus   & 6.0 \cite{RemageEvans:2006pp} & 9.66 \cite{Wenger:2000sw} & Dynamical & galactic \\ \hline

  Polaris   & 4.5 \cite{Evans:2008hq} & 3.9715 \cite{Wenger:2000sw} & Dynamical & galactic \\ \hline

OGLE-GD-CEP-1884   & 7.47 \cite{2024ApJ...965L..17S} & 78.14 \cite{Wenger:2000sw} & Pulsation model & galactic \\ \hline

  V1334 Cyg   & 4.288 \cite{2018ApJ...867..121G} & 3.331 \cite{Wenger:2000sw} & Dynamical & galactic \\ \hline
   
\end{tabular}
\end{table}

The existence of the Cepheids S Muscae (S Mus) \cite{RemageEvans:2006pp} and OGLE-GD-CEP-1884 \cite{2024ApJ...965L..17S} with dynamically and pulsation determined masses, respectively,  serve as two potential benchmark observations. We note that Cepheids with dynamically determined masses are inherently more conservative with respect to constraining axions. The reason is that axions are expected to alter both stellar evolution models (as we have shown in this work) as well as stellar pulsation models (which is the subject of future work by the current authors~\cite{futurework}). Dynamical mass determination, conducted through measurements of binary orbital parameters, is immune to axion physics.\footnote{In this context, we note in passing the Cepheid mass discrepancy problem \cite{2023arXiv230712386G}:  while Cepheid masses determined from pulsation models  and binary orbital dynamics are largely in concordance with each other,  they are both  different from masses expected from stellar evolution. Axions are unlikely to alleviate this problem, since the effect of axions appears to be most relevant to the radius of the core and hence the core potential and the triggering of the blue loop.} S Mus, with a dynamically determined mass, thus serves as our most conservative benchmark.

\subsection{S Mus: Luminosity and the Cepheid Mass Discrepancy Problem}

\begin{figure}[t]
  \centering
    \includegraphics[width=0.63\textwidth]{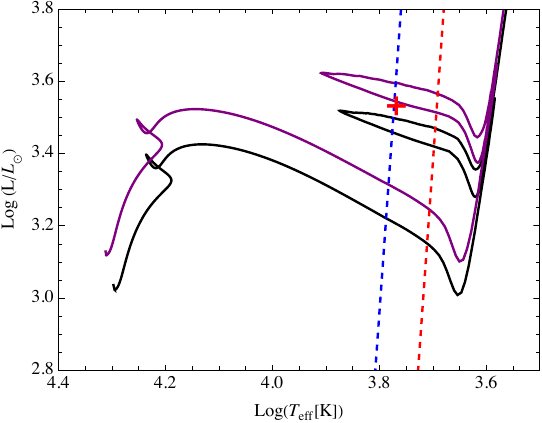}
   \caption{Reproducing the observed luminosity of the Cepheid S Mus with a $6 M_\odot$ model (black) and a $6.4 M_\odot$ model (purple). The metallicity is set to $Z = 0.014$ and the overshoot to $\alpha_{\rm ov} = 0.1$ for both models. Axions are absent ($g_{10} = 0 $). The red dot signifies the luminosity of S Mus $\log L/L_{\odot} = 3.54$ following the analysis of observational data performed by \cite{2013AJ....146...93E}, while for the effective temperature we have taken the value quoted in \cite{1997ApJ...477..916B}.}
   \label{fig:SMus}
\end{figure}

Utilizing observational data from Cepheids to constrain axions immediately runs into a complication. While dynamical mass determination fixes the mass of S Mus as shown in Table~\ref{tab:cepheidsobs},  taking this same mass as a benchmark in evolutionary models leads to a  Cepheid with lower luminosity than required to match data. This is the well-known mass discrepancy problem, already described in the Introduction. We illustrate this problem in the case of S Mus in Fig.~\ref{fig:SMus} with a $6 M_\odot$ model (black) and a $6.4 M_\odot$ model (purple). The metallicity is set to $Z = 0.014$ and the overshoot to $\alpha_{\rm ov} = 0.1$ for both models. Axions are absent ($g_{10} = 0 $). The red dot signifies the observed luminosity of S Mus $\log L/L_{\odot} = 3.54$, according to the analysis of observational data performed by \cite{2013AJ....146...93E}. It should be noted that the observational collaboration did not report uncertainties in the luminosity. which therefore could not be incorporated in our study. On the other hand, the benchmark effective temperature is taken to be 5850 K, following the determination  by \cite{1997ApJ...477..916B}. These authors took a pulsation period of 9.6 days, a pulsation mass of $6 M_\odot$, and interpolated evolution models with luminosities between  $\log L/L_{\odot} = 3.50 \,-\,3.75 $. While uncertainties in the temperature also undoubtedly exist, we do not incorporate them in our benchmark depiction in Fig.~\ref{fig:SMus}, since they have not been reported in the aforementioned study.

It is clear that a $6 M_\odot$ model is unable to reach the required luminosity to match with observation, even after incorporating $\alpha_{\rm ov} = 0.1$, while a $6.4 M_\odot$ model does so. Moreover, these conclusions would change as $\alpha_{\rm ov}$ is varied, and also with variations in the exact position of S Mus in the instability strip: the second crossing, versus the third, and so on, which is also unknown. Therefore, the situation depicted in Fig.~\ref{fig:SMus} is a benchmark among a variety of possible evolutionary tracks. We report  these possibilities in Table~\ref{tab:SMusnoaxion}, varying the mass and overshoot.

\begin{table}[h]
\centering
\caption{S Mus: The mass and overshoot required to achieve $\log L/L_{\odot} = 3.54$ and effective temperature 5850 K in the HR diagram (red dot in Fig.~\ref{fig:SMus}). We have assumed that $g_{10} = 0$. }
\label{tab:SMusnoaxion}
\vspace{3pt}
 \begin{tabular}{|c | c | } 
 \hline\hline
 Mass & $\alpha_{\rm ov}$  \\ \hline
 
 $6.0 \, M_\odot$   & 0.2 \\ \hline
 $6.4 \, M_\odot$   & 0.1 \\ \hline
 $6.6 \, M_\odot$   & 0.0 \\ \hline

\end{tabular}
\end{table}

\begin{table}[h]
\centering
\caption{S Mus: 
The values of $g_{10}$ for which the blue loop just disappears up to the red edge of the instability strip are reported, for the stellar properties listed in Table~\ref{tab:SMusnoaxion}. }
\label{tab:SMusyesaxion}
\vspace{3pt}
 \begin{tabular}{|c | c | c | } 
 \hline\hline
 Mass & $\alpha_{\rm ov}$ & $g_{10}$  \\ \hline
 
 $6.0 \, M_\odot$   & 0.2  & 1.6 \\ \hline
 $6.4 \, M_\odot$   & 0.1  & 1.7 \\ \hline
 $6.6 \, M_\odot$   & 0.0  & 0.35 \\ \hline

\end{tabular}
\end{table}

\begin{figure}[h]
  \centering
    \includegraphics[width=0.65\textwidth]{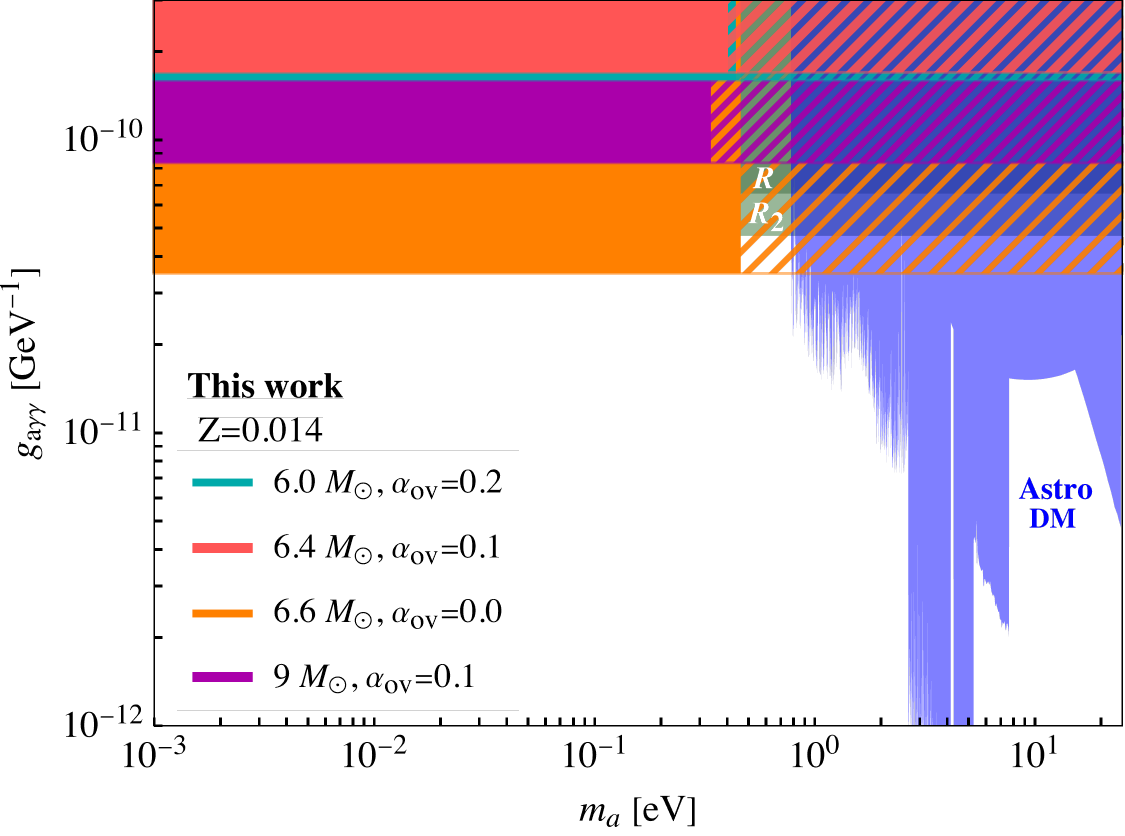}
  \caption{The constraints on the plane of $(g_{a\gamma\gamma}, m_a)$ for a fixed value of  $Z = 0.014$, for three stellar models corresponding to the observed luminosity of S Mus:  $6 M_{\odot}$ with $\alpha_{\rm ov} = 0.2$ (green), $6.4 M_{\odot}$ with $\alpha_{\rm ov} = 0.1$ (red) and $6.6 M_{\odot}$ with $\alpha_{\rm ov} = 0.0$ (orange). These benchmark values are taken from Table~\ref{tab:SMusyesaxion}. The results for a $9  M_{\odot}$ model with $\alpha_{\rm ov} = 0.1$ (purple) are also displayed.  The constraints on the coupling are obtained from the violation of Eq.~\ref{corepotcrit}. The range of $m_a$ depicted by hatched regions corresponds to values smaller than the star’s lowest core temperature (where the majority of axion production occurs), while the range shown by shaded regions is smaller than the lowest temperature across the entire stellar profile. We also show existing constraints from the stellar $R$-parameter (green)~\cite{Ayala:2014pea} and $R_2$-parameter (light green)~\cite{Dolan:2022kul}, as well as astrophysical axion dark matter searches (blue, ``Astro DM'')~\cite{Janish:2023kvi,Yin:2024lla,Todarello:2023hdk,Grin:2006aw,Nakayama:2022jza,Carenza:2023qxh,Porras-Bedmar:2024uql}. The constraints are obtained from~\cite{AxionLimits}.  
  }
  \label{fig:Sensitivity}
\end{figure}

Switching on axion emission and requiring the blue loop to vanish would lead to different constraints on  $g_{10}$, depending on  the mass of the model, the value of $\alpha_{\rm ov}$, and the position of S Mus in the instability strip. Of these options, we keep the position of S Mus fixed, and report the constraints on $g_{10}$ as the overshoot and mass are varied in Table~\ref{tab:SMusyesaxion}. The procedure we have followed is as follows. The mass and overshoot for a given model that achieves the position of S Mus on the HR diagram in the absence of axions is first taken from Table~\ref{tab:SMusnoaxion}, and then the value of $g_{10}$ that just nullifies the loop up to the red edge of the instability strip  for that mass and overshoot is reported. The corresponding constraints on the plane of $(g_{a\gamma\gamma}, m_a)$ coming from S Mus are depicted  in Fig.~\ref{fig:Sensitivity}. 

The results for a  $9 M_{\odot}$ model, which  is the heaviest model that was simulated in the current work, are also shown. In the absence of simulations of blue loops at higher stellar masses, and hence information about the effect of axions on such blue loops, we instead attempt to deduce such information from the linear fit in Fig.~\ref{fig:varyingmassconstraints}. We  obtain
\be \label{linift}
g_{10} = (-0.28 \pm 0.02) \, \frac{M}{M_\odot}\, + \,(3.5 \pm 0.1) \,\,. 
\ee
We show the constraints corresponding to the largest model we can simulate, the $9 M_\odot$ model corresponding to Fig.~\ref{fig:9msunconstraints}, in purple regions in Fig.~\ref{fig:Sensitivity}. 

The fact that we display the results for a $6.6 M_\odot$ model may be open to criticism, given that it is outside of the range for the allowed dynamical mass of S Mus, which only goes up to $6.4 M_\odot$.  Moreover, the fact that we first determine the mass and overshoot to match the observed luminosity of S Mus, and \textit{then} vary $g_{10}$ until the blue loop vanishes, is not the most precise method to set limits. More correctly, what should be done is to scan the three-dimensional parameter space comprising $\{M, \alpha_{\rm ov}, g_{10}\}$ and select the subspace where the observed luminosity of S Mus is achieved \textit{and} the blue loop vanishes; the values of $g_{10}$ from that subspace would furnish the constraints on the axion coupling. While we do not pursue this method, our results indicate that in the large mass - small overshoot regime ($\sim 6.6 M_\odot$) the constraints on the axion are likely to become stringent. This further underscores our main point: that heavy, luminous Cepheids are a fertile target for these kinds of studies.

\subsection{Incorporating Overshoot Uncertainties for $6 M_\odot$ and $8 M_\odot$ Models}\label{overshootuncert}

In this Section, we investigate how our results change with changes in overshoot. 

The most important physical quantity influencing the development and morphology of the blue loop is the size of the core. The dependence of the core radius on the composition for a fixed core mass has been studied by various authors \cite{10.1093/mnras/156.2.129, 10.1093/mnras/stu2666} and generally while composition is important, it is not the dominant factor in determining the blue loop. The main determinant in the development of the blue loop is the outward motion of the hydrogen burning shell that is evident in the progressive panels labelled ``A'', ``Instability Strip'', and ``E'' in Fig.~\ref{fig:hydprofile} \cite{Lauterborn:1971, 10.1093/mnras/156.2.129, 10.1093/mnras/stu2666}. This outward motion reduces the amount of excess helium above the shell, which is a critical requirement for the blue loop to occur.

The position of the discontinuous jump in the hydrogen profile in panel ``A'' of Fig.~\ref{fig:hydprofile} marks the innermost point of convective penetration (the first dredge-up). The further inwards (smaller values of $m_{\rm enc}$) this penetration occurs in stage ``A'', the more robust the ensuing blue loop. Conversely, the inward penetration of the convective envelope is limited for stars with mass above $9 M_\odot$ at solar metallicity, inhibiting the blue loop in such cases \cite{10.1093/mnras/stu2666}. The  efficiency of convective mixing in the core \cite{2019ApJ...886...27W, 2020ApJ...894..118W} and the efficiency below the convective envelope \cite{1991A&A...244...95A, 2009MNRAS.396.1833G} play an important role in determining the size of the convective core.

Models using the Ledoux criterion require overshooting to match observations \cite{1991A&AS...89..451M}, since overshooting extends the core beyond the convective boundary predicted by  stellar theory. Convective cells in the stellar core penetrate into the radiative regions due to their non-negligible velocity, an effect parametrized by $\alpha_{\rm ov}$. Larger overshoot results in larger stellar cores and higher luminosities, as well as higher stellar radii. Constraining $\alpha_{\rm ov}$ has been a focus of activity, with efforts ranging from using eclipsing binary stars ($\alpha_{\rm ov}$ between 0.25 and 0.32 for stars between 2.5 and 7 $M_\odot$) \cite{1997MNRAS.285..696S} to asteroseismological measurements ($\alpha_{\rm ov}  = 0.44 \pm 0.07$ for $\theta$ Ophiuchi) \cite{10.1111/j.1365-2966.2007.12142.x} and non-rotating evolutionary models ($\alpha_{\rm ov}$ between 0.25 and 0.30) \cite{1986A&A...158...45M}. Based on VLT Flames survey of massive stars ranging from $5-60 M_{\odot}$, \cite{brott:2011} found $\alpha_{\rm ov} = 0.335$. In Sec.~\ref{sec: blueloopaxion}, we followed the initial conditions of \cite{Ekstrom:2012}  who constrained overshoot using the main sequence width for mass ranges between $1.35-9 M_{\odot}$. The overshoot was set to $\alpha_{\rm ov} = 0.1$ during hydrogen and helium burning for masses above 1.7 $M_\odot$, making it applicable for our case. We now show how our results change as $\alpha_{\rm ov}$ is varied.

\begin{figure}[t]
  \centering
    \includegraphics[width=0.6\textwidth]{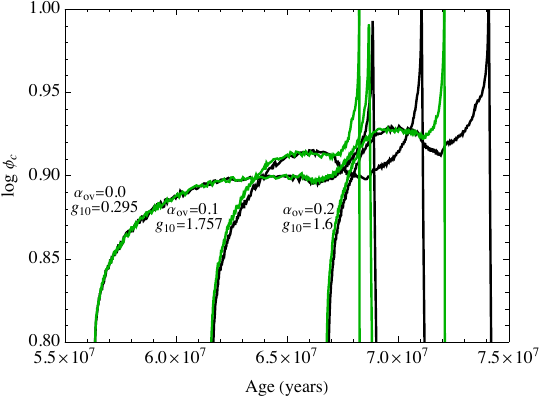}
  \caption{The behavior of the core potential for $\alpha_{\rm ov} = 0.0, 0.1,$ and $0.2$ with $g_{10} = 0$ (black) and the value of $g_{10}$ that just eliminates the blue loop in each case (green). The results depicted here are for a $6 M_{\odot}$ model. The metallicity is set to $Z = 0.014$.}
  \label{fig:overshoot6M}
\end{figure}

\begin{figure}[b]
  \centering
    \includegraphics[width=0.6\textwidth]{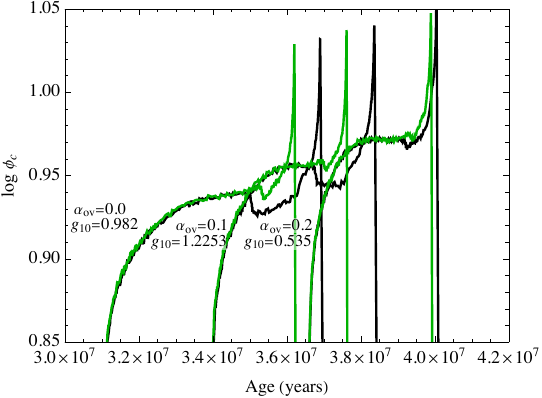}
  \caption{The behavior of the core potential for $\alpha_{\rm ov} = 0.0$, $0.1$, and $0.2$ with $g_{10} = 0$ (black) and the value of $g_{10}$ that just eliminates the blue loop in each case (green). The results depicted here are for a $8 M_{\odot}$ model. The metallicity is set to $Z = 0.014$.}
  \label{fig:overshoot8M}
\end{figure}

We show the behavior of the core potential for $\alpha_{\rm ov} = 0.0, 0.1,$ and $0.2$ for a $6 M_\odot$ model in Fig.~\ref{fig:overshoot6M} and for a $8 M_\odot$ model in Fig.~\ref{fig:overshoot8M}, respectively. The black curves show the case of $g_{10} = 0$, while the green curves show the core potential for values of $g_{10}$ at which the blue loop is just curtailed at the red edge of the instability strip. There are two features that are important in Fig.~\ref{fig:overshoot6M}. Firstly, as  $\alpha_{\rm ov}$ becomes larger, the critical value of the core potential $\phi_{\rm crit}$ increases. This is because as the overshoot increases,  the core becomes enlarged both in mass as well as in radius. Secondly, as $\alpha_{\rm ov}$ increases, the core potential displays a more pronounced dip when it falls below $\phi_{\rm crit}$ and thus the blue loop is more robust; it therefore requires a larger value of $g_{10}$ to attenuate it. On the other hand, for smaller  $\alpha_{\rm ov}$, the core potential becomes flatter and just barely dips below the corresponding value of $\phi_{\rm crit}$; even a small value of $g_{10}$ is sufficient to disrupt the blue loop in this case. We note that the constraints on $g_{10}$ do not change appreciably for larger values of $\alpha_{\rm ov}$ in the $6 M_\odot$ model. Ultimately, both the $6 M_\odot$ model as well as the $8 M_\odot$ model lose their blue loops for large enough overshoot. This is because the large overshoot increases the mass of the core to an extent that the core potential acquires a  large value and no longer dips below $\phi_{\rm crit}$. It should be noted that the overshoot physics of an $8 M_{\odot}$ model can differ substantially from that of a  $6 M_{\odot}$ model, as axions are switched on. Thus, in Fig.~\ref{fig:overshoot8M}, it is clear that the constraint coming from  $\alpha_{\rm ov} = 0.2$ is substantially stronger than that coming from  $\alpha_{\rm ov} = 0.1$, and even  $\alpha_{\rm ov} = 0.0$.

We pause to warn the reader against a possible misperception emanating from Figs.~\ref{fig:overshoot6M} and \ref{fig:overshoot8M}. The core potential should not be misconstrued as a quantity with predictive power: indeed, the fact that $\phi_c$ before the loop increases
with increasing overshooting does not predict that a higher $g_{10}$
value is needed to suppress the loop. Similarly,  after the onset of the loop, $\phi_c$ tends to increase again, although the loop is still in its blue-sided part
(and vice versa). A more correct way to think of the core potential is as a diagnostic for the onset of a loop.

\begin{figure}[ht]
  \centering
    \includegraphics[width=0.49\textwidth]{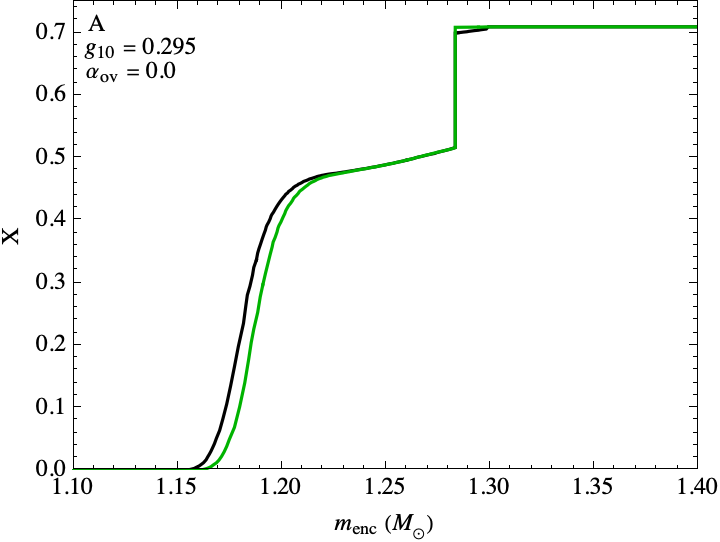} \includegraphics[width=0.49\textwidth]{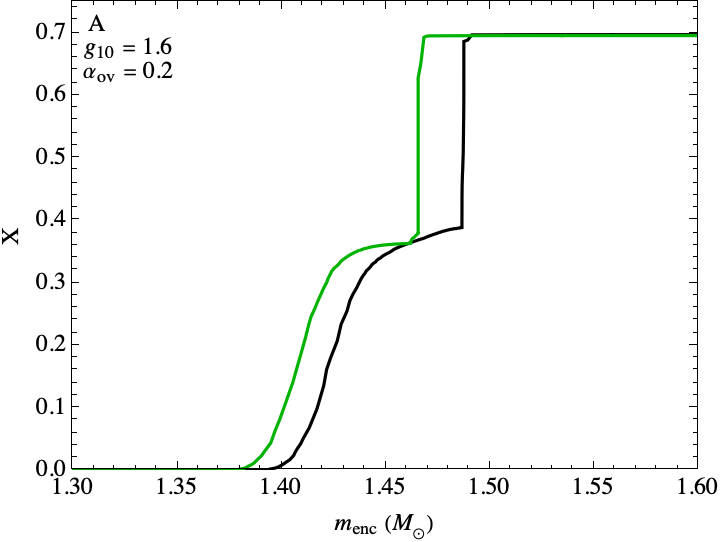}\\
    \includegraphics[width=0.49\textwidth]{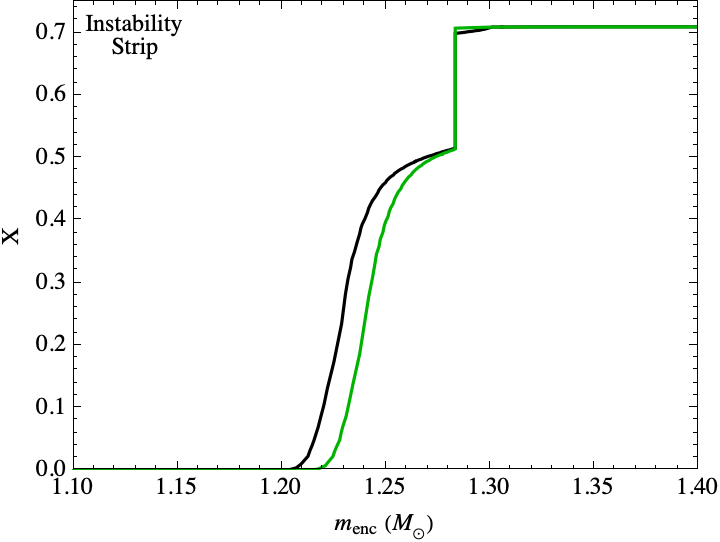} \includegraphics[width=0.49\textwidth]{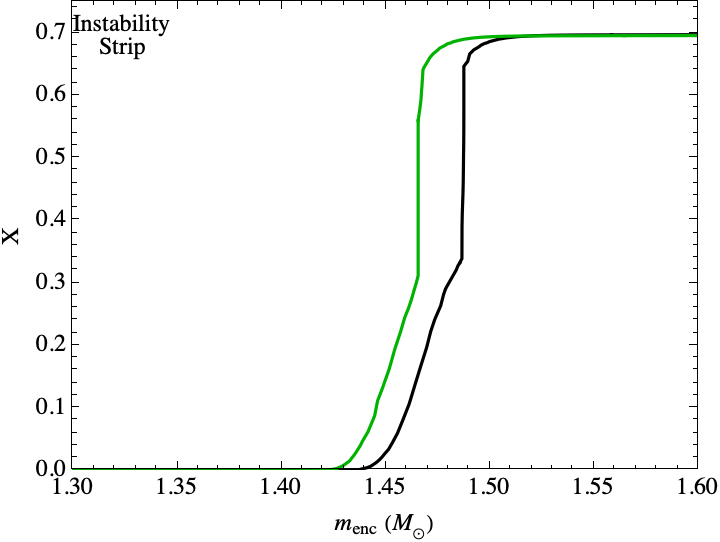}\\
    \includegraphics[width=0.49\textwidth]{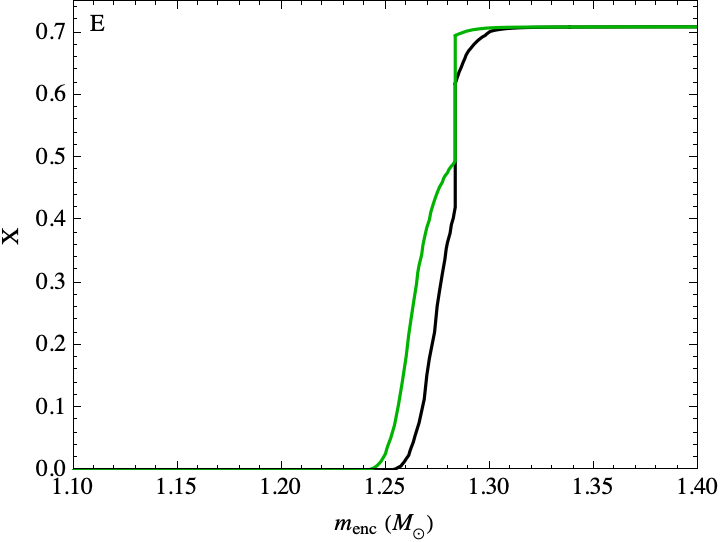} \includegraphics[width=0.49\textwidth]{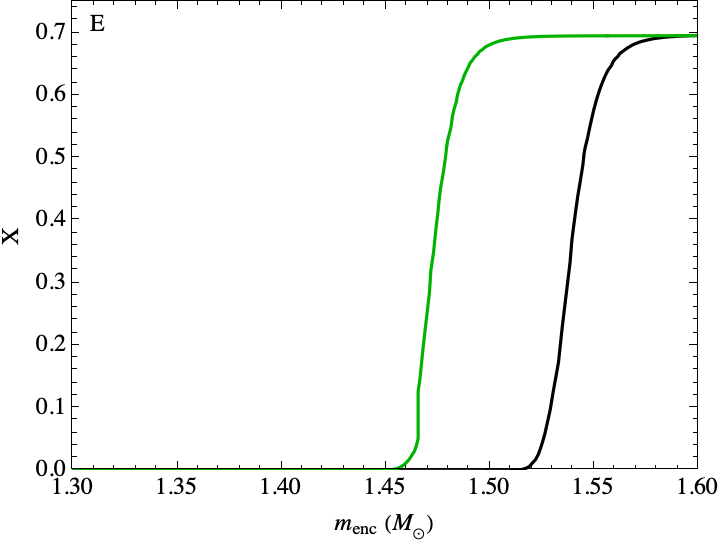}
  \caption{The hydrogen mass fraction $X$  profile of a 6 $M_\odot$ model before (top panels), during (middle panels) and after (bottom panels) the blue loop stage. The axion coupling is set to $g_{10} = 0 $ for the black curves. Left column: overshoot $\alpha_{\rm ov} = 0.0$ and $g_{10} = 0.295 $ (green). Right column: overshoot $\alpha_{\rm ov} = 0.2$ and $g_{10} = 1.6 $ (green). The metallicity is set to $Z = 0.014$.
  }
  \label{fig:hydprofilealpha0002}
\end{figure}

In Fig.~\ref{fig:hydprofilealpha0002}, we display the hydrogen mass fraction of a 6 $M_\odot$ model with $\alpha_{\rm ov} = 0.0$ (left) and $\alpha_{\rm ov} = 0.2$ (right), respectively, for $g_{10} = 0$ and the value of $g_{10}$ that just switches off the blue loop in each case. For each column, the top panel shows the profile at stage ``A'', the middle panel shows the profile at a point in the instability strip, while the bottom panel shows the profile at stage ``E''. The first feature to note is that  in the absence of axions (black curves), as $\alpha_{\rm ov}$ becomes larger, the location of the first dredge-up moves outwards, enlarging the core and creating a more robust blue loop. For stage ``A'', the location of the jump in the hydrogen profile occurs at $m_{\rm enc} = 1.28, \, 1.35,$ and  $1.49$ for $\alpha_{\rm ov} = 0.0, 0.1,$  and 0.2, respectively, as can be seen from  Fig.~\ref{fig:hydprofilealpha0002} and Fig.~\ref{fig:hydprofile}. The effect of axions is to shift the hydrogen profiles, especially the location of the first dredge-up, to the left. This is especially evident in the cases of $\alpha_{\rm ov} = 0.1, 0.2$ where the jump in the profile is relocated to 1.30 (a reduction of 3.7\%) and 1.47 (a reduction of 1.3\%), respectively. The case of $\alpha_{\rm ov} = 0.0$ shows a barely perceptible left-ward shift of the profile at the first dredge-up, showing how sensitive the system is to changes in the core. The location of the hydrogen burning shell at stage ``A'' for the $\alpha_{\rm ov} = 0.0$ model is at slightly larger enclosed mass for $g_{10} = 0.295$ compared to $g_{10} = 0$, although by stage ``E'' the situation is reversed.

\begin{figure}[t]
  \centering
    \includegraphics[width=0.49\textwidth]{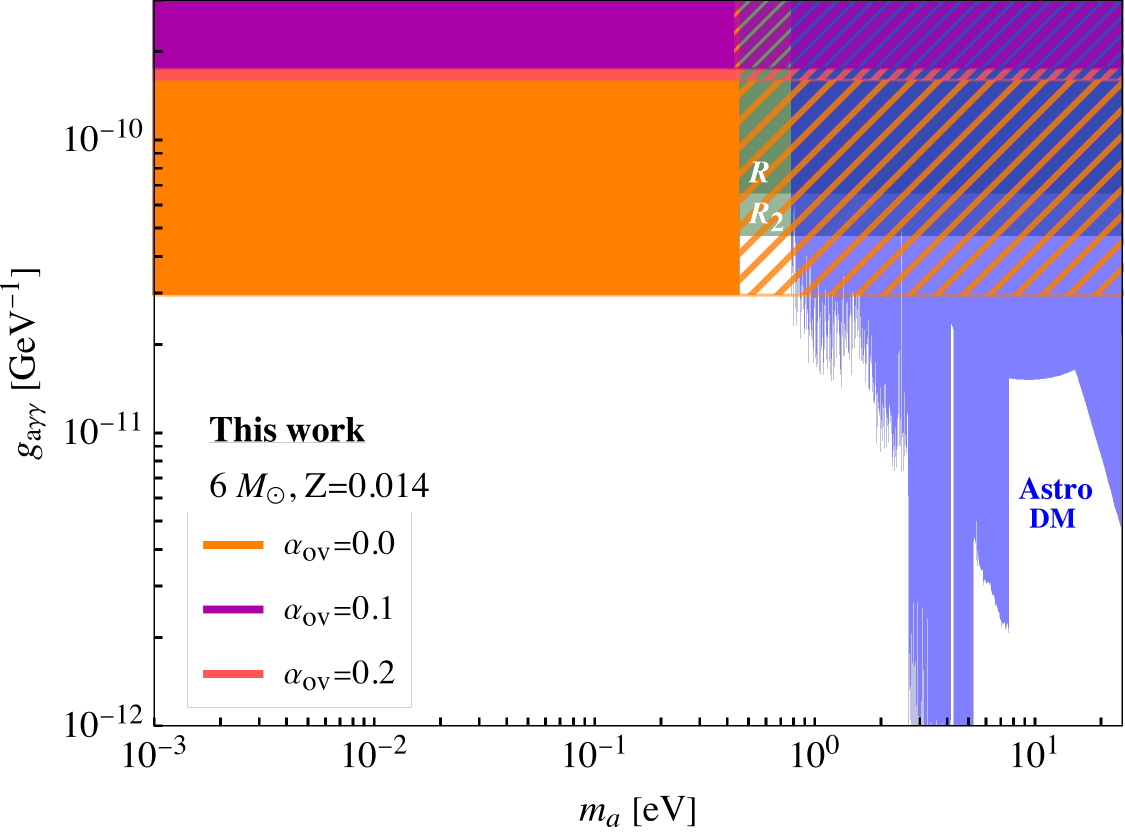}\hspace{5pt}     \includegraphics[width=0.49\textwidth]{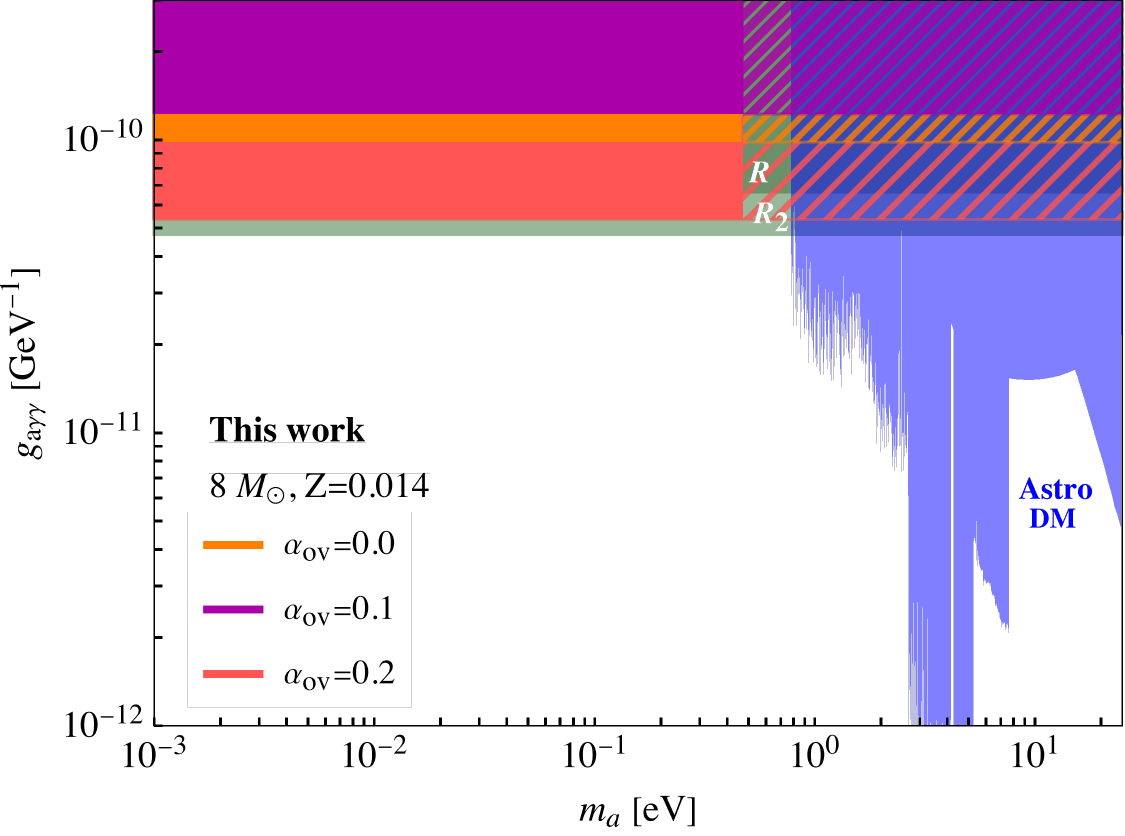}
  \caption{Left panel: The constraints on the plane of $(g_{a\gamma\gamma}, m_a)$ for a $6 M_{\odot}$ model with $\alpha_{\rm ov} = 0.0$ (orange), $\alpha_{\rm ov} = 0.1$ (purple), and $\alpha_{\rm ov} = 0.2$ (red). Right panel: Corresponding constraints for a $8 M_{\odot}$ model. The constraints on the coupling  are obtained from the violation of Eq.~\ref{corepotcrit}. The metallicity is set to $Z = 0.014$. }
  \label{fig:overshootsenivitivty}
\end{figure}

In Fig.~\ref{fig:overshootsenivitivty}, we depict constraints on the plane of $(g_{a\gamma\gamma}, m_a)$ for two benchmark models:  a $6 M_{\odot}$ model (left panel)  and $8 M_{\odot}$ model (right panel), and for each model we vary values of $\alpha_{\rm ov}=0.0$ (orange), $\alpha_{\rm ov}=0.1$ (purple), and $\alpha_{\rm ov}=0.2$ (red). As discussed above, the blue loop phase becomes more robust with an increase in overshoot from $0.0$ to $0.1$, requiring a larger $g_{a\gamma\gamma}$ coupling to completely eliminate the phase. Conversely, the blue loop phase becomes unstable when $\alpha_{\rm ov}\sim0.2$, as can be seen from the black curves in Fig.~\ref{fig:overshoot6M} and Fig.~\ref{fig:overshoot8M}. This leads to an enhanced sensitivity in the red-shaded regions of Fig.~\ref{fig:overshootsenivitivty}, especially in the case of the $8M_\odot$ model.

\section{Conclusions}
\label{sec: conclusion}

In this paper, we have explored the effect of axion emission on the development of the blue loop stage of intermediate mass stars, taking the core potential  
as a diagnostic for the onset of the blue loop. As we have stressed in Sec.~\ref{overshootuncert}, the amount of convective overshooting drastically changes the critical value of the axion coupling for which loops are suppressed; moreover, the core potential also depends on the detailed chemical composition (as a function of the radius) of the star, as depicted in Fig.~\ref{fig:hydprofile}. Rotation, metallicity, and nuclear reaction rates also affect the core potential. The critical value of the core potential, $\phi_{\rm crit}$, is determined from simulations and thus $\phi_{\rm crit}$ can only be determined from the models, i.e., once the loop sets in. In particular, it should be understood that $\phi_{\rm crit}$ has no predictive power and can be used only as a diagnostic or even proxy for the onset of a loop.

Assuming galactic metallicity and $\alpha_{\rm ov} = 0.1$, the existence of the Cepheid S Mus with dynamically determined mass $6 M_\odot$ has been used to  constrain the axion-photon coupling. The effect of different choices of the overshoot on the underlying physics has been studied. Results have also been provided for $9 M_\odot$ models, which constitute the limits of our simulation. Prospects and challenges for obtaining reliable bounds from the simulation of  heavier Cepheids have been discussed, and tentative projected bounds for a $12 M_\odot$ model is presented in Appendix~\ref{12solarmass}. Although the blue loop stage of more massive stars is more robust in the absence of energy loss via axion emission, our findings suggest that, for a given overshoot, axions have a greater impact on more massive stars by eliminating their blue loops due to their higher core temperatures. Consequently, observational evidence from more massive Cepheid stars could impose stronger constraints on the $g_{a\gamma\gamma}$ coupling.

Compared to the previous study of axion effect on Cepheids~\cite{Friedland:2012hj}, our study included axion energy loss from both degenerate medium and non-degenerate medium in the simulation for a more precise description of axions originating from the dense helium core, whereas the previous study only took into account the non-degenerate medium contribution. Further improvements can be included in future works to reduce the uncertainty in the simulation of Cepheids. Firstly, we used the variable $\zeta$ in Eq.~\ref{eq:zetaelectron} to classify stellar regions with and without degenerate electron contributions. A refined approach could be taken when a more sophisticated model of the transition region around the hydrogen-burning shell is available. Regarding the charge screening effect due to high charge density, we treated the electron screening and the ion screening independently since their correlation wavenumbers differ significantly. Future studies could explore the correlations in the emission regions near the core boundary. Finally, the uncertainty in calculating the emission rate can be reduced by adopting a finer resolution for the cell thickness in \texttt{MESA} simulations, as the $\epsilon_a \propto T^7$ dependence on the stellar temperature profile makes the calculation highly sensitive to even small variations in $T$. 

Further future directions have been described at various points in the main text. Axions may influence the location of the instability strip, potentially leading to observable consequences. The effect of axions on the PL relations of Cepheids is a very interesting direction to pursue, especially since the performance of Cepheids as standard candles hinges on percent-level control over systematic uncertainties, which may very well be challenged by the emission of axions. The incorporation of rotation and different values of the metallicity, and their interplay with axion physics, is another interesting future direction.  We remark that in our simulations axions more readily eliminate the blue loop phase of heavier stars when assuming larger $\alpha_{\rm ov}$ values, while lighter stars are more easily affected under null overshooting. This calls for more comprehensive overshoot models to refine future axion constraints from Cepheids.

One of the goals of our paper has been to motivate the simulation community to further investigate the prospects of simulating blue loops  for heavier stars with overshoot, given the importance of the heavier stars for fundamental physics that we have emphasized. Of course, it is also possible that current simulations are adequate and Cepheids above $9 M_{\odot}$ do not exist. 

Blue loops have acquired a reputation, in our opinion undeserved, of being disfavored for setting constraints on axions, due to uncertainties in the microphysics and numerical description \cite{DiLuzio:2021ysg}. Such an evaluation may be premature, given the advances in numerical simulations and evolution and pulsation models on the one hand, and the anticipated discovery of dynamically mass-determined Cepheids from Gaia on the other. \textit{Our main message is that the observation, dynamical mass determination, and reliable blue loop simulation of heavy Cepheids is likely to yield strong constrains on the axion-photon coupling.}

\acknowledgments
We thank Sean Matt for discussions  and Georg G. Raffelt for very helpful correspondence. We also thank Frederick Hiskens and Raymond Volkas for helpful correspondence. K.A. was supported by the REU program at the University of Oklahoma.  P.S.~is supported in part by NSF grant PHY-2412834.  The research activities of K.S. and T.X. are supported in part by the U.S. National Science Foundation under Grant PHY-2412671. 

\appendix

\section{Constraints from $12 M_\odot$ Models: Prospects and Challenges}
\label{12solarmass}

Given the fact that constraints on axions are projected to be very powerful for a $12 M_\odot$ model, we make a few brief comments about the prospects of converting such bounds into reality. 

\begin{figure}[h]
  \centering
    \includegraphics[width=0.65\textwidth]{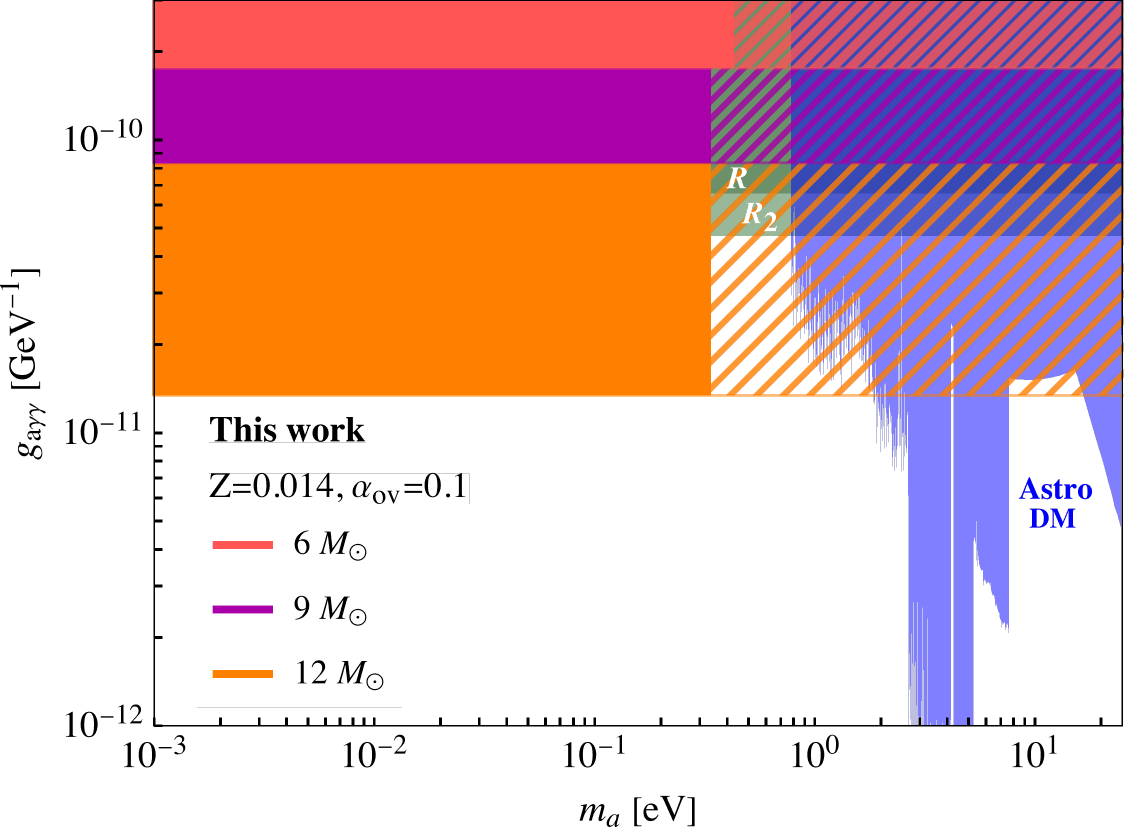}
  \caption{The constraints on the plane of $(g_{a\gamma\gamma}, m_a)$ for a fixed value of $\alpha_{\rm ov} = 0.1$, $Z = 0.014$, and three stellar models: $6 M_{\odot}$ (red), $9 M_{\odot}$ (purple), and $12 M_{\odot}$ (orange). The constraints on the coupling are obtained from the violation of Eq.~\ref{corepotcrit}. The range of $m_a$ depicted by hatched regions corresponds to values smaller than the star’s lowest core temperature (where the majority of axion production occurs), while the range shown by shaded regions is smaller than the lowest temperature across the entire stellar profile. The constraints from the $12 M_{\odot}$ model are obtained from Eq.~\ref{linift} and correspond neither to an actual observed candidate with dynamically determined mass, nor to a simulation. We also show existing constraints from the stellar $R$-parameter (green)~\cite{Ayala:2014pea} and $R_2$-parameter (light green)~\cite{Dolan:2022kul}, as well as astrophysical axion dark matter searches (blue, ``Astro DM'')~\cite{Janish:2023kvi,Yin:2024lla,Todarello:2023hdk,Grin:2006aw,Nakayama:2022jza,Carenza:2023qxh,Porras-Bedmar:2024uql}. The constraints are obtained from~\cite{AxionLimits}. }
\label{fig:Sensitivity12}
\end{figure}

Firstly, although Cepheids of such mass have not been confirmed through dynamical methods, there are tentative indications that such candidates may have appeared in Gaia data, with  masses determined by pulsation models. In particular, we comment on the treatment of \cite{2020ApJ...898L...7M}, where  nonlinear convective pulsation models were developed for a wide selection of stellar input parameters, and then applied to a sample of the Gaia DR2 Galactic Cepheids database. Mass-dependent Period-Wesenheit relations in the Gaia bands were derived, which enables mass-dependent estimates of individual distances, and hence, after matching with astrometric distances, individual masses of Cepheids. The DR2 sample peaked at $5.6 M_\odot$ and $5.4 M_\odot$ for variables pulsating in the fundamental and first overtone modes, respectively. After parallax offset, the peaks moved to  $5.2 M_\odot$ and $5.1 M_\odot$, respectively. In both cases, at least one candidate with mass $12 M_\odot$ was obtained, as can be seen from Fig. 2 of \cite{2020ApJ...898L...7M}. These trends broadly match earlier studies of Galactic Cepheids using Gaia DR2, for example that of \cite{2019A&A...623A.117K}. Of course, it is important to issue the appropriate caveats here. We reiterate that the mass estimate by \cite{2020ApJ...898L...7M} is not conclusive since it is based on 1-d pulsation models. Similarly, the "calibration" with the results from \cite{2019A&A...623A.117K} is done for much lower stellar masses. Moreover, neither the models predict loops, as we discuss below.

If one were to take the observation of a $12 M_\odot$ Cepheid as a given, there are still two obstacles to obtaining a trustworthy bound on axions. The first, and most important one, is that we were unable to simulate a blue loop for such a model in \texttt{MESA} even in the absence of an axion, let alone study the effects of axion emission in such models. For the $12 M_\odot$ model, the inability to obtain a blue loop persists even for $\alpha_{\rm ov} = 0.0$. For a $10 M_\odot$ or $11 M_\odot$ model, on the other hand, blue loops cannot be obtained for  $\alpha_{\rm ov} = 0.1$, although they can be obtained for $\alpha_{\rm ov} = 0.0$. As we noted previously, this appears to be a general feature of current stellar evolution codes, including the Cambridge STARS code \cite{10.1093/mnras/stu2666}. At this conjuncture, we are unable to resolve this problem. The second obstacle is that the pulsation models may themselves be affected by axion emission. This obstacle offers a promising route to study axions that instead relies on their effects on pulsation, rather than their effects in curtailing the blue loop itself. This is left for future work.

The projections for a $12 \, M_\odot$ model are depicted in Fig.~\ref{fig:Sensitivity12}. We utilize the linear fit in Fig.~\ref{fig:varyingmassconstraints} and Eq.~\ref{linift}. The $12 M_\odot$ model constraints should be considered with the appropriate caveat that they are obtained neither from simulation nor from observation; moreover, we cannot justify the linear fit in Eq.~\ref{linift} at a theoretical level. However, Eq.~\ref{linift} and the limits coming from the $12 M_\odot$ model do underscore the general physics point: \textit{the observation and mass determination of heavier Cepheids would correspond to stronger constraints on axions.}

\section{General Comments on Post-Blue Loop Evolution} \label{postblueloop}

In this section, we make a few preliminary comments about the possible effects of axions on the post-blue loop evolution of stars. A full treatment is kept for future work.

The first comment pertains to Fig. \ref{fig:Radius}, which shows the evolution of the stellar and core radius versus age. The production of axions not only  reduces the maximum size of the core radius, but also ushers in the AGB stage sooner in the evolutionary history. This is clear by the fact that the onset of the AGB occurs at $\sim 7.2 \times 10^7$ years, $\sim 7.1 \times 10^7$ years, and $\sim 6.9 \times 10^7$ years, for $g_{10} = 0$ (black), $g_{10} = 1$ (red), and $g_{10} = 2$ (green), respectively. This confirms that the helium burning stage is shortened with the inclusion of axion loss, a fact also discussed in \cite{2017A&A...605A.106C}.

\begin{figure}[h]
  \centering
    \includegraphics[width=0.65\textwidth]{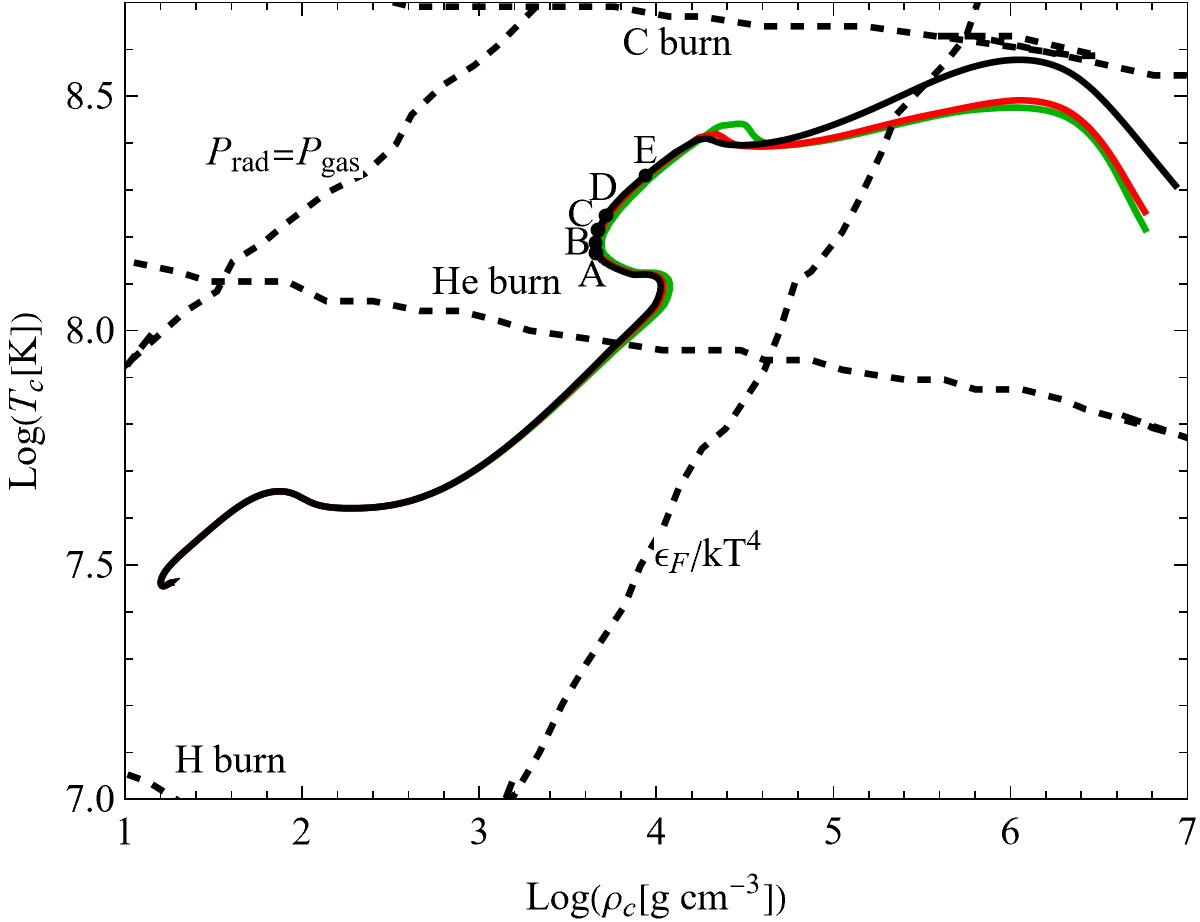}
  \caption{Log central temperature vs Log central density for a 6$M_\odot$ model. Burning bounds are extrapolated from \texttt{MESA}'s pgstar plots.}
  \label{fig:RhoT}
\end{figure}

\begin{figure}[h]
  \centering
    \includegraphics[width=0.65\textwidth]{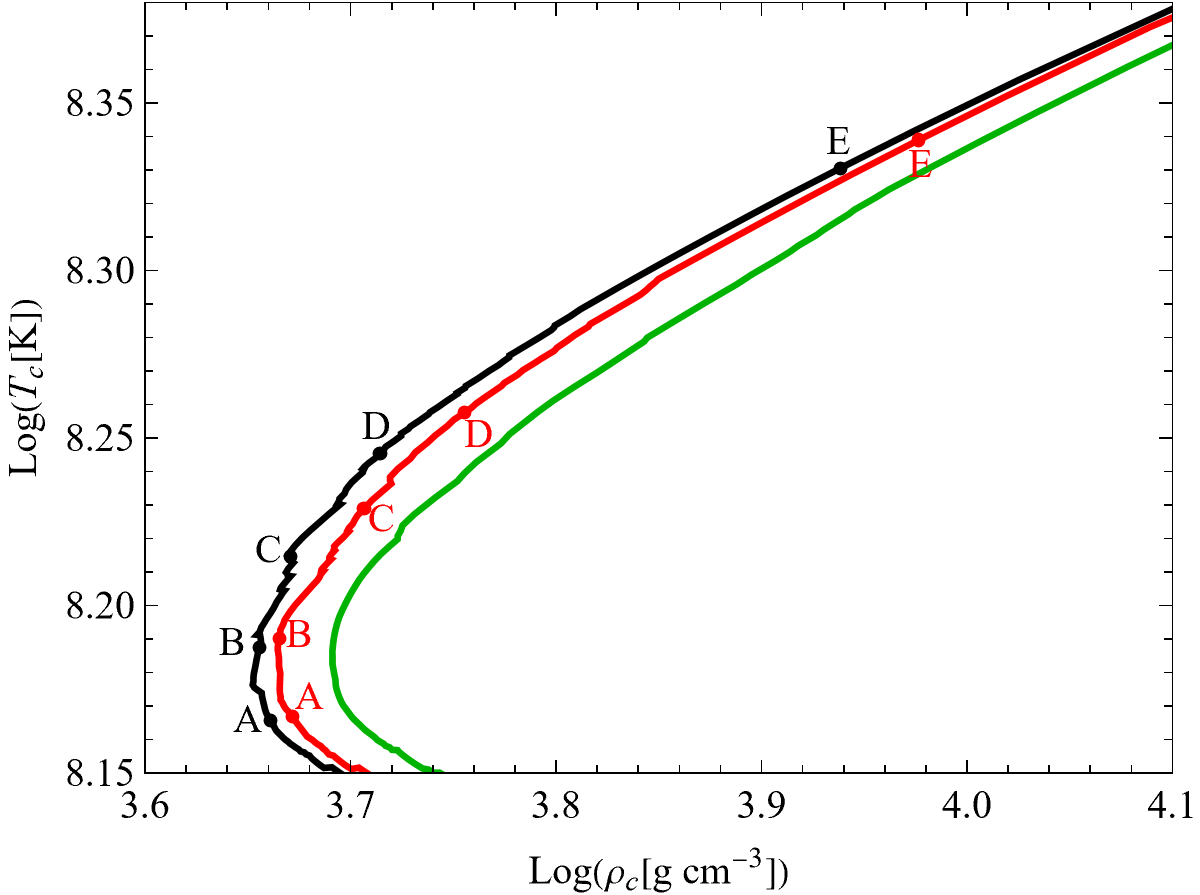}
  \caption{Zoomed in of Fig.~\ref{fig:RhoT}. Symbols are discussed in the text.  }
  \label{fig:RhoTzoomed}
\end{figure}

In Fig. \ref{fig:RhoT}, we depict the evolutionary trajectory of a 6$M_\odot$ model on the plane of $\log T_{c}$ vs $\log \rho_{c}$. Fig.~\ref{fig:RhoTzoomed} shows the same evolution, zoomed in on the blue loop stage, and displaying the various stages ``A'' through ``E'' described before. It is clear from the zoomed in figure that an increase in $g_{10}$ lowers $T_{c}$ and increases $\rho_{c}$ during the blue loop stage. The effect is exacerbated during the AGB stage of evolution, as  depicted by the deviation from the black line in Fig. \ref{fig:RhoT}. A decrease in luminosity from axion emission near the core leads to a decrease in  the temperature.

\bibliographystyle{JHEP}
\bibliography{ref}

\end{document}